\documentclass[12pt]{article}
\pdfoutput=1
\usepackage{tikz}
\usetikzlibrary{matrix,arrows,decorations.pathmorphing,decorations.pathreplacing}
\usepackage{jheppub}
\usepackage{amsmath,amssymb,euscript,array,mathrsfs,appendix,ctable,marvosym}
\usepackage{arydshln}
\usepackage{todonotes}
\usepackage{graphicx}
\usepackage[normalem]{ulem}

\newcommand\blank[1]{#1}
\renewcommand\blank[1]{}
\def\Buildrel#1\over#2\under#3{\mathrel{\mathop{\kern0pt
#2}\limits^{#1}_{#3}}}

\newcommand{\otoprule}{\midrule[\heavyrulewidth]}

\usepackage{blkarray}

\def\LL{{\mathscr L}}
\def\JI{J}
\def\JN{{\EuScript J}}
\def\CF{{\cal F}}

\def\DG{d_{\mathfrak g}}
\def\DH{d_{\mathfrak h}}
\def\DF{d_{\mathfrak f}}
\def\DFf{d_{{\mathfrak f}^{(1)}}}
\def\JJ{\mathscr{J}}
\def\AA{\EuScript A}

\def\mpsu{\mathfrak{psu}}

\def\SO{\text{SO}}

\def\mg{\mathfrak g}
\def\mf{\mathfrak f}
\def\mh{\mathfrak h}

\newcommand{\Tr}{\operatorname{Tr}}
\newcommand{\STr}{\operatorname{STr}}

\newcommand{\Ad}{\operatorname{ad}}
\newcommand{\AD}{\operatorname{Ad}}

\def\B0{{\boldsymbol 0}}

\def\SU{\text{SU}}

\def\Dbarslash{\,\,{\raise.15ex\hbox{/}\mkern-12mu {\bar D}}}
\def\Dslash{\,\,{\raise.15ex\hbox{/}\mkern-12mu D}}
\def\delslash{\,\,{\raise.15ex\hbox{/}\mkern-9mu \partial}}
\def\delbarslash{\,\,{\raise.15ex\hbox{/}\mkern-9mu {\bar\partial}}}

\def\AA{{\EuScript A}}

\newcommand{\EQ}[1]{\begin{equation}\begin{split} #1
\end{split}\end{equation}}

\newcommand{\FIG}[1]{\begin{figure}[ht]\begin{center} #1 \end{center}\end{figure}}

\title{Integrable Deformations of Strings on Symmetric Spaces}
\author[a]{Timothy J. Hollowood,}
\author[b]{J. Luis Miramontes}
\author[a,c]{and David M. Schmidtt}
\affiliation[a]{Department of Physics, Swansea University, Swansea, SA2 8PP, U.K.}
\affiliation[b]{Departamento de F\'\i sica de Part\'\i culas and IGFAE,
Universidad de Santiago de Compostela, 15782 Santiago de Compostela, Spain}
\affiliation[c]{Instituto de F\'\i sica Te\'orica IFT/UNESP, Rua Dr. Bento Teobaldo Ferraz 271, Bloco II, CEP 01140-070, Sa\~o Paulo-SP, Brasil}

\emailAdd{t.hollowood@swansea.ac.uk}
\emailAdd{jluis.miramontes@usc.es}
\emailAdd{david.schmidtt@gmail.com}

\abstract{A general class of deformations of integrable sigma-models with symmetric space $F/G$ target-spaces are found. These deformations involve defining the non-abelian T dual of the sigma-model and then replacing the coupling of the Lagrange multiplier imposing flatness with a gauged $F/F$ WZW model. The original sigma-model is obtained in the limit of large level. The resulting deformed theories are shown to preserve both integrability and the equations-of-motion, but involve a deformation of the symplectic structure. It is shown that this deformed symplectic structure involves a linear combination of the original Poisson bracket and a generalization of the Faddeev-Reshetikhin Poisson bracket which we show can be re-expressed as two decoupled $F$ current algebras. It is then shown that the deformation can be incorporated into the classical model of strings on ${\mathbb R}\times F/G$ via a generalization of the Pohlmeyer reduction. In this case, in the limit of large sigma-model coupling it is shown that the theory becomes the relativistic symmetric space sine-Gordon theory. These results point to the existence of a deformation of this kind for the full Green-Schwarz superstring on AdS$_5\times S^5$.}

\notoc
\begin{document}

\maketitle

\newpage

\pgfdeclarelayer{top layer}
\pgfdeclarelayer{foreground layer}
\pgfdeclarelayer{background layer}
\pgfsetlayers{background layer,main,foreground layer,top layer}
\section{Introduction}

One of the most fascinating underlying features of the original AdS/CFT correspondence is its integrability (see the reviews \cite{Beisert:2010jr,Arutyunov:2009ga}). 
In particular, the world-sheet theory for the string moving in $AdS_5\times S^5$ in the Green-Schwarz formalism is a sigma-model whose target-space is a quotient of a Lie super-group by an ordinary Lie group known as a semi-symmetric space \cite{Serg} (see also \cite{Zarembo:2010sg}). In the context of AdS$_5\times S^5$, the semi-symmetric space is $\text{PSU}(2,2|4)/\text{Sp}(2,2)\times\text{Sp}(4)$ where the numerator reflects the superconformal symmetry of the dual ${\cal N}=4$ gauge theory. The bosonic sector of the semi-symmetric space can written as a product of two ordinary symmetric spaces $\SO(1,5)/\SO(1,4)\times\SO(6)/\SO(5)$, that is AdS$_5\times S^5$ itself.

It is clearly interesting to investigate deformations of the sigma-model which preserve integrability, and whether these more general theories can be interpreted as strings moving in deformations of the AdS$_5\times S^5$ background. Questions of this type have been tackled in various different ways. 
One approach, that is completely quantum, is to move focus from the string background and to look for deformations of the exact $S$-matrix that describes excitations on the string world-sheet around a certain classical background. The $S$-matrix for these excitations has a Yangian symmetry structure associated to the unbroken 
symmetry which takes the form of two copies of the triply extended super Lie algebra $\mh=\mpsu(2|2)\ltimes{\mathbb R}^3$. One way to deform this $S$-matrix is very natural 
from the theory of quantum integrable field theories; namely, 
to deform the symmetry into the quantum group $U_q(\mh)$. Here, $q$ is the deformation parameter with the original structure given in the limit $q\to1$. The S-matrix of the deformed theory is built on the $R$-matrix of the quantum group written down in \cite{Beisert:2008tw} and can be shown to satisfy all the axioms of $S$-matrix theory as long as either $q$ is real and arbitrary or $q=e^{i\pi/k}$ for an integer $k$ \cite{Hoare:2011wr,Hoare:2012fc}. In the former case the $S$-matrix appears in the vertex form whereas in the latter it involves the interaction-round-a-face (IRF) or restritced-solid-on-solid (RSOS) form \cite{Hoare:2013ysa}. This is crucial in order to satisfy hermitian analyticity and unitarity.
The restriction alluded to here follows from the fact that the deformation parameter of the quantum group is a root of unity. This kind of restriction has been studied in integrable field theories in the past  \cite{LeClair:1989wy,Bernard:1990cw,Bernard:1990ys}.
The Thermodynamic Bethe Ansatz of the $q$-deformed model has been investigated in \cite{Arutyunov:2012zt,Arutyunov:2012ai}.

The case of $q$ real was studied from a sigma-model perspective in \cite{Delduc:2013fga}. It involves deforming the target-space background of the Green-Schwarz sigma-model \cite{Delduc:2013qra,Delduc:2014kha,Kawaguchi:2014qwa}. This kind of deformed sigma-model is a generalization of the so-called Yang-Baxter deformation introduced by Klim\v{c}\'\i k \cite{Klimcik:2002zj}. 
The explicit background fields for this deformed theory, called the $\eta$ deformation, were found in \cite{Arutyunov:2013ega}. Other work involving this or related deformations appears in \cite{Arutynov:2014ota,Arutyunov:2014cda,Hoare:2014pna,Kameyama:2014bua,Kawaguchi:2014fca}.

In the present work we focus on the $q$ deformed theory with $q$ a root of unity. This case is not covered by the Yang-Baxter type sigma-model deformation of \cite{Delduc:2013fga,Delduc:2013qra,Delduc:2014kha}. The deformation is more subtle because $k$ is quantized and the limit $k\to\infty$ 
for which the original theory is---at least na\"\i vely---recovered is not a continuous limit. It is known that the $q$ root-of-unit deformation shares a common structure with the real $q$ case in that it can be described as a deformation of the symplectic structure of the underlying theory \cite{Delduc:2013fga,Delduc:2013qra,Hollowood:2014fha}. What is missing in the $q=e^{i\pi/k}$ case is a Lagrangian formulation of the deformed theory and it is the main goal of the present work to fill this gap. 

Generally, the string sigma-model and its deformation, once gauge is fixed, breaks two-dimensional Lorentz symmetry. It was a fascinating observation  \cite{Pohlmeyer:1975nb,Grigoriev:2007bu,Mikhailov:2007xr,Grigoriev:2008jq} that the equation-of-motion of the sigma-model could be written as a relativistic equation, which initiated a detailed study of the resulting so-called generalized sine-Gordon theories at the quantum level \cite{Hollowood:2011fq,Hoare:2011nd,Hollowood:2013oca}.
It was guessed that these relativistic theories should be obtained from a deformation of the sigma-model. In the present work this picture is finally made precise. In particular, we show that when the sigma-model coupling becomes large for fixed $k$, the deformed theory becomes precisely the relativistic generalized sine-Gordon theory.

In the present work, we will lay the ground work for an investigation of the $q=e^{i\pi/k}$ deformed string theory by focussing on the bosonic sector of such theories. To this end, we consider a
new class of deformations of integrable bosonic sigma-models with ordinary symmetric space target-spaces $F/G$. The deformations we find are completely different from those associated to the Yang-Baxter sigma-model found by Delduc et al.~in \cite{Delduc:2013fga}; however, they share the same deformed Poisson structure albeit related by a simple analytic continuation. 

The paper is organized as follows. In section \ref{s2} and \ref{s3} we focus on integrable deformations of the principal chiral model which are a useful warm up for the symmetric space sigma-models. We first recall some old work on deformations of the $\SU(2)$ principal chiral model by Balog et al.~\cite{Balog:1993es} and at the quantum level in \cite{Evans:1994hi}. We then describe how Sfetsos \cite{Sfetsos:2013wia} generalized these kind of integrability preserving deformations to arbitrary groups. The key feature here is that the deformation can be formulated at the Lagrangian level as a deformed WZW model. We show that the underlying integrability can be understood in a very simple way in terms of this WZW formulation. In section \ref{s3}, we investigate the Hamiltonian structure of this deformed WZW model and show that the constrained Dirac brackets consist of two commuting classical current, or Kac-Moody, algebras. These brackets are related by an analytic continuation in $q$ to those constructed in \cite{Delduc:2013fga} relevant to the $q$ real case.

In section \ref{s4} we turn to the main focus of the work. We show that the formalism applied to the principal chiral model can be suitably modified to apply to symmetric space sigma-models. In particular, we show how integrability can be maintained in the deformed theory. In section \ref{s5}
we turn to the Hamiltonian structure of the deformed theories. The analysis here is more complicated than the principal chiral model as there are a mixture of first and second class constraints that must be disentangled. We show how the Poisson brackets derived by Delduc et al.~\cite{Delduc:2013fga} for the Yang-Baxter type deformation of the symmetric space sigma-models are related to those of our deformation by a simple analytic continuation.

Section \ref{s6} is devoted to studying string motion on a symmetric space target-space after gauge fixing. This involves imposing the Virasoro constraints in a process known as Pohlmeyer reduction \cite{Pohlmeyer:1975nb} reviewed in depth in \cite{Miramontes:2008wt}. The additional constraints must be implemented on the phase space and lead to a new Dirac bracket. In section \ref{s7} we show that in the limit of large sigma-model coupling, at fixed $k$, this new Poisson structure is precisely that of a generalized sine-Gordon theory.

The final section addresses some issues about the deformed theories at the quantum level as well as setting out how the picture we have established for the bosonic sector of the string theory can be lifted to the complete superstring theory including all the fermionic fields.

\section{Deformed Principal Chiral Model}\label{s2}

In this section, we will investigate the simpler case of the principal chiral model drawing on existing work in the literature to provide inspiration for the symmetric space case considered in section \ref{s4}. In particular, we turn to the work of Balog, Forg\'acs, Horv\'agh and Palla in \cite{Balog:1993es}. The starting point is the principal chiral model whose action can be written\footnote{We take 2d metric $\eta_{\mu\nu}=\text{diag}(1,-1)$. We often use the null coordinates $x^\pm=t\pm x$ and for vectors we have $A^\pm=A^0\pm A^1$ and $A_\pm=(A_0\pm A_1)/2$ so that the invariant $A_\mu B^\mu=2(A_+B_-+A_-B_+)$.}
\EQ{
S_\text{PCM}=\frac{\kappa^2}{4\pi}\int d^2x\Tr\big(\partial_\mu f^{-1}\partial^\mu f\big)\ ,
}
for a field $f(x,t)$ valued in a Lie group $F$.\footnote{We take a basis of anti-hermitian generators $T^a$ for the Lie algebra $\mathfrak f$ of $F$ with $[T^a,T^b]=f^{abc}T^c$. We will take the normalisation $\text{Tr}(T^aT^b)=-\delta^{ab}$ in the defining representation.}

If we define the current $\JI_\mu=f^{-1}\partial_\mu f$, then the equations-of-motion are simply the conservation condition
\EQ{
\partial_+ \JI_-+\partial_-\JI_+=0\ .
\label{em1}
}
In addition, it follows from its definition that the current satisfies the  Cartan-Maurer identity
\EQ{
\partial_+\JI_--\partial_- \JI_++[\JI_+,\JI_-]=0\ .
\label{cm1}
}
Taken together, these two equations can be written as a Lax equation, that is the flatness condition
\EQ{
[\partial_++{\mathscr L}_+(z),\partial_-+{\mathscr L}_-(z)]=0
\label{leq2}
}
for a $z$-dependent gauge field
\EQ{
{\mathscr L}_\pm(z)=\frac{z}{z\pm1}\JI_\pm
}
where $z$, the spectral parameter, is an arbitrary parameter. It is well known that the existence of a Lax connection implies that the field theory admits an infinite number of non-local conserved quantities and renders the theory classically integrable \cite{Luscher:1977rq,Brezin:1979am,Bernard:1990jw}. For example, the next simplest conserved current is defined by
\EQ{
\JN_\pm=\pm\JI_\pm+\frac12[\JI_\pm,\phi]\ ,\qquad \partial_\pm\phi=\mp\JI_\pm\ ,
}
which is non-local because of the definition of $\phi$ in terms of $\JI_\mu$.

However, the same equations-of-motion for the current $\JI_\mu$ also follow from a completely different theory by writing instead $\JI_\mu=\epsilon_{\mu\nu}\partial^\nu\Phi$ for a field $\Phi$ valued in the Lie algebra $\mathfrak f$ of $F$ with action
\EQ{
S'=\frac{\kappa^2}{4\pi}\int d^2x\Tr\big(-\partial_\mu\Phi\partial^\mu\Phi+\frac13\epsilon^{\mu\nu}[\Phi,\partial_\mu\Phi]\partial_\nu\Phi\big)\ .
}
Some properties of this theory are investigated by Nappi in \cite{Nappi:1979ig} who showed that, although $S'$ and $S_\text{PCM}$ have the same equations-of-motion, when written in terms of the current $\JI_\mu$, the are physically inequivalent having different renormalization group properties for instance.

In fact, Rajeev \cite{Rajeev:1988hq} had already pointed that the Poisson structure of the principal chiral model admits a one-parameter deformation of the form
\EQ{
&\{J_\pm^a(x), J_\pm^b(y)\}_{\hat{x}}= \frac{\pi}{2\kappa^2} f^{abc}(J_\mp^c -(1+2\hat{x}) J_\pm^c)\delta(x-y) \pm \frac{\pi}{\kappa^2} \delta^{ab}\delta'(x-y)\,,\\[5pt]
&\{J_+^a(x), J_-^b(y)\}_{\hat{x}}=-\frac{\pi}{2\kappa^2} f^{abc}(J_+^c +J_-^c) \delta(x-y)
\label{gt1}
}
for a parameter $\hat x$. Then, the $\hat x$-independent equations-of-motion~\eqref{em1} and \eqref{cm1} can be written as
\EQ{
\partial_t J_\pm^a= \{J_\pm^a, H_{\text{PCM}}\}_{\hat{x}}\,, \qquad
H_{\text{PCM}}=-\frac{\kappa^2}{2\pi}\int dx\, \Tr(J_+^2 + J_-^2)\,.
}
The important point is that these Poisson brackets satisfy the Jacobi identity for any value of $\hat x$. Setting $\hat x=1$ we recover the Poisson brackets of the principal chiral model, whereas setting $\hat x=-1$ gives the Poisson brackets of the theory defined by the action $S'$.

Rajeev \cite{Rajeev:1988hq} established the remarkable fact that the deformed Poisson brackets \eqref{gt1} can be decoupled into two commuting classical Kac-Moody algebras with currents
\EQ{
\widetilde{\JJ}_\pm=-\frac{\kappa^2 {\cal N}}{2\pi}\left\{\left(\frac{1}{\hat{x}+1} + \frac{1}{\sqrt{\hat{x}^2-1}} \right) J_\pm 
+\left(\frac{1}{\hat{x}+1} - \frac{1}{\sqrt{\hat{x}^2-1}} \right) J_\mp
\right\}\,,
\label{ksq}
}
obeying
\EQ{
\big\{\widetilde{\JJ}^a_\pm(x),\widetilde{\JJ}^b_\pm(y)\big\}_{\hat{x}}&={\cal N}\left(f^{abc}\widetilde{\JJ}_\pm^c(y)\delta(x-y)\pm\frac {\hat{k}{\cal N}}{2\pi}\delta^{ab}\delta'(x-y)\right)\ ,\\
\big\{\widetilde{\JJ}^a_+(x),\widetilde{\JJ}^b_-(y)\big\}_{\hat{x}}&=0\ ,\\
\label{km3}
}
where
\EQ{
\hat{k}=\frac{2\kappa^2}{(\hat x+1)\sqrt{\hat x^2-1}}\ ,
\label{cch}
}
and we have introduced an overall normalization factor ${\cal N}$ that will be fixed later.
It is important to recognise that the Kac-Moody currents are not chirally conserved as in a WZW model. In addition, it is worth pointing out that the fact that the central terms appear with signs $\pm$ is entirely conventional of classical Kac-Moody algebras, as opposed to their quantum cousins; for example, see Bowock's classic paper on the Hamiltonian structure of (gauged) WZW models \cite{Bowcock:1988xr}.

What was missing from Rajeev's original work was a simple Lagrangian formulation of the deformed theories. This gap was filled by 
Balog et al.~\cite{Balog:1993es} for the case of $F=\SU(2)$, who showed that the deformed theories, for general $\hat x$, could be derived from a conventional sigma-model with torsion
\EQ{
S=\frac{\kappa^2}{4\pi}\int d^2x\Big[G_{ab}(X)\partial_\mu X^a\partial^\mu X^b-B_{ab}(X)\epsilon^{\mu\nu}\partial_\mu X^a\partial_\nu X^b\Big]\ .
\label{act1}
}
The explicit expressions for the metric $G$ and $B$-field are complicated and will not be needed. However, what is significant for our narrative is that there are 
two distinct branches of theories depending on whether $\hat x^2\gtrless1$, denoted as case b), c), and d) in \cite{Balog:1993es}.\footnote{There 
is another branch of solutions denoted as a) in \cite{Balog:1993es} that will play no r\^ole in our discussion.} 

The branch with $\hat x^2\leq1$ naturally interpolates between the principle chiral model ($\hat x=1$) and the theory $S'$ ($\hat x=-1$). The branch with $\hat x^2>1$ is qualitatively different. On this branch one of the coordinates is most naturally interpreted as an angle and, consequently, the $B$-field term in \eqref{act1} naturally takes the form of a Wess-Zumino term and there is a quantization of the parameter 
$k={\cal N}\hat{k}$. This is natural, because it is only on this branch that $k$, interpreted as the central charge of the classical Kac-Moody algebras in \eqref{cch}, is real. 
On this branch, the action \eqref{act1} in the limit $\hat x\to\infty$ with $\kappa^2\to\infty$ such that $\hat{k}$ is fixed
is precisely the $\text{SU}(2)$ WZW model action.

The deformed Poisson brackets \eqref{gt1} can also be written
\EQ{
&\{J_0^a(x),J_0^b(y)\}_{\hat{x}}= -\frac{\pi}{\kappa^2}(\hat{x}+1) f^{abc} J_0^c \delta(x-y)\,,\\[5pt]
&\{J_1^a(x),J_1^b(y)\}_{\hat{x}}= -\frac{\pi}{\kappa^2}(\hat{x}-1) f^{abc} J_0^c \delta(x-y)\,,\\[5pt]
&\{J_0^a(x),J_1^b(y)\}_{\hat{x}}= -\frac{\pi}{\kappa^2}(\hat{x}+1) f^{abc} J_1^c \delta(x-y) + \frac{2\pi}{\kappa^2} \delta^{ab}\delta'(x-y)\,.
}
These are precisely the deformed Poisson brackets of the principle chiral model described in by Delduc et al.~\cite{Delduc:2013fga}, up to suitable convention-depend re-scalings, with the deformation parameter defined in that reference equal to
\EQ{
\epsilon^2=\frac{1-\hat x}2\ .
}
In fact \cite{Delduc:2013fga} identifies the deformed Poisson brackets above as a linear combination of the Poisson brackets of the principal chiral model and the Faddeev-Reshetikhin model \cite{Faddeev:1985qu}:
\EQ{
\{F,G\}_{\hat{x}}=\{F,G\}_\text{PCM}+\epsilon^2\{F,G\}_\text{FR}\ .
}
Importantly, the deformation considered in \cite{Delduc:2013fga} corresponds to $\epsilon^2\geq0$ and it turns out to be defined only for $0\leq\epsilon^2<1$; i.e., it corresponds to the branch $\hat x^2\leq 1$ (more precisely, to $-1< \hat x\leq1$). In the present paper we will be interested in the branch $\hat{x}\geq1$, which corresponds to analytically continuing to $\epsilon^2\leq0$ and is not covered by the analysis in \cite{Delduc:2013fga}.\footnote{Notice that $\hat{x}\leq-1$ corresponds to $\epsilon^2\geq1$.}

The $\SU(2)$ theories of Balog et al.~\cite{Balog:1993es} on the $\hat x^2\geq1$ branch were considered at the quantum level in \cite{Evans:1994hi}.  We will have more to say about the resulting quantum picture in section \ref{s8}, however, it is 
worth pointing out even at this stage that $k=\hat{k} {\cal N}$ appears as a central term in the classical Kac-Moody algebras \eqref{km3} and it should not be surprising that in the quantum theory $k$ will be quantized in the standard way as a positive integer. Therefore, the deformations on the $\hat x^2\geq1$ branch are actually discrete in nature and the principal chiral model is obtained by a discrete limiting procedure as $k\to\infty$.

\subsection{Extending beyond $\SU(2)$}

As it stands, the analysis of \cite{Balog:1993es} turned out difficult to extend beyond $\text{SU}(2)$ because of the complexity of the explicit formulae. Recently progress was made on this problem by Sfetsos \cite{Sfetsos:2013wia} who writes down a remarkably simple Lagrangian formulation of the deformed theories on the branch $\hat x\geq1$. This formulation
involves starting with the original principal chiral model for an $F$-valued field $f$ and then adding to it the WZW model for an $F$-valued field $\CF $. One then gauges a common $F$ symmetry corresponding to
\EQ{
f\to Uf\ ,\qquad \CF =U\CF U^{-1}\ ,
\label{fact}
}
for $U\in F$. 

It is important that the WZW field is gauged with vector action in order to avoid a gauge anomaly. 
So the action of the deformed theory is the sum
\EQ{
S[f,\CF ,A_\mu]=S_\text{gPCM}[f,A_\mu]+S_{\text{gWZW}}[{\cal F},A_\mu]\ .
}
The gauging of the PCM involves simply replacing derivatives by covariant derivatives:
\EQ{
S_\text{gPCM}[f,A_\mu]=-\frac{\kappa^2}{\pi}\int d^2x\Tr\Big[f^{-1}(\partial_+ f+A_+ f)f^{-1}(\partial_- f+A_- f)\Big]\ ,
}
while the gauged WZW action is \cite{Karabali:1988au,Gawedzki:1988hq,Karabali:1989dk}\footnote{$k\in\mathbb Z>0$ for a unitary group $F$, while $k\in\frac12\mathbb Z>0$ for sympletic or orthogonal groups.}
\EQ{
S_\text{gWZW}[\CF ,A_\mu]&=-\frac k{2\pi}\int d^2x\Tr\Big[
\CF ^{-1}\partial_+\CF \,\CF ^{-1}\partial_-\CF +2A_+\partial_-\CF \CF ^{-1}\\ &~~~~~~~~~
-2A_-\CF ^{-1}\partial_+\CF -2\CF ^{-1}A_+\CF  A_-+2A_+A_-\Big]
\\ &~~~~~~~~~+\frac k{12\pi}\int d^3x\,\epsilon^{abc}\Tr\,\Big[\CF ^{-1}\partial_a\CF \,
\CF ^{-1}\partial_b\CF \,\CF ^{-1}\partial_c\CF \Big]\ .
\label{gWZW}
}

The gauge symmetry in the resulting theory may be simply fixed by imposing the condition that 
$f=I$, the identity. The nice feature of this gauge fixing is that the resulting theory can then be viewed as a deformation of the $F/F$ gauged WZW theory with action 
\EQ{
S_\text{def}[\CF ,A_\mu]=S_\text{gWZW}[\CF ,A_\mu]-\frac k{\pi}\Big(\frac1{\lambda}-1\Big)\int d^2x\,\Tr\,\big(
A_+A_-\big)\ ,
\label{dWZW}
}
where we have defined the deformation parameter
\EQ{
\lambda=\frac{k}{k+\kappa^2}\ .
\label{ldf}
}
This parameter is related to $\hat x$ introduced earlier and $\epsilon^2$ of \cite{Delduc:2013fga} via
\EQ{
\hat x=\frac{1+\lambda^2}{2\lambda}\ ,\qquad\epsilon^2=-\frac{(1-\lambda)^2}{4\lambda}\ .
}
On the branch of interest in this work $0\leq\lambda\leq1$, which corresponds to $\hat x\geq 1$.

It is important to remark that we have already fixed the gauge symmetry, so the resulting theory \eqref{dWZW} is {\it not\/} invariant under gauge transformations for generic $\lambda$. The theory at $\lambda=1$ is the undeformed theory which is gauge invariant and so it is perhaps understandable that the limit $\lambda\to1$, i.e.~$\hat x\to 1$, must be considered carefully.

The formulation of the deformed theory as a deformation of a gauged $F/F$ WZW model is undeniably elegant and, moreover, it lies in the class of generalized WZW theories investigated by Tseytlin in \cite{Tseytlin:1993hm}.\footnote{In order to compare, Tseytlin defined the matrix $Q=(2-\lambda^{-1})I$.}

In \cite{Sfetsos:2013wia} Sfetsos shows that the equations-of-motion of \eqref{dWZW}  can be written as \eqref{em1} and \eqref{cm1} with a suitable definition of the current $\JI_\mu$. The idea is to first eliminate the fields $A_\pm$ by using their equations-of-motion and then define the current via
\EQ{
\JI_\mu=\frac2{1+\lambda}A_\mu\ ,
\label{gww}}
which yields\footnote{Here, we have define the adjoint action by a group element $\AD(\CF)x=\CF x\CF^{-1}$ in the defining representation. For an algebra element, we define the adjoint action $\Ad(a)x=[a,x]$.}
\EQ{
\JI_+&=\frac{2\lambda }{1+\lambda }\big(\AD(\CF)-\lambda \big)^{-1}\partial_+\CF \CF ^{-1}\ ,\\
\JI_-&=-\frac{2\lambda }{1+\lambda }\big(\AD(\CF^{-1})-\lambda \big)^{-1}\CF ^{-1}\partial_-\CF \ .
\label{icr}
}
These components satisfy
\EQ{
\partial_\mp \JI_\pm=\pm\frac12[\JI_+,\JI_-]\ .
\label{h45}
}
which are equivalent to \eqref{em1} and \eqref{cm1}. Sfetsos goes on to show that 
the suitably normalized Poisson brackets take the form \eqref{gt1} using the sigma-model formulation of the deformed theories. In section~\ref{s3} we will obtain the same result in a simpler way by starting directly from their formulation as deformed gauge WZW theories. 

\subsection{Non-abelian T-duality and bosonized Thirring model}

It is worth pausing to explain one of the principal motives behind \cite{Sfetsos:2013wia} and also the earlier \cite{Sfetsos:1994vz,Polychronakos:2010hd}. There is an equivalent way to formulate the principal chiral model as a first order system. In this formulation, one takes an $\mathfrak f$-valued gauge field $A_\mu$ as a fundamental field and then the fact that ultimately there is a group-valued field $f$ for which $A_\mu=f^{-1}\partial_\mu f$ is imposed by means of an $\mathfrak f$-valued Lagrange multiplier $\nu$. Note that previously we denoted this quantity as the current $\JI_\mu$ but here, since its r\^ole is rather different we call it $A_\mu$. The action of the system is
\EQ{
S_\text{PCM}=-\frac{\kappa^2}{\pi}\int d^2x\Tr\big(A_+A_-+\nu F_{+-}\big)
\label{apcm}
}
where 
\EQ{
F_{+-}=\partial_+A_--\partial_-A_++[A_+,A_-]\ ,
}
is the single non-vanishing component of the curvature of $A_\mu$. The vanishing of $F_{+-}$ means that $A_\mu$ is pure gauge and that implicitly there exists a group valued field such that $A_\mu=f^{-1}\partial_\mu f$.

The formulation above is naturally the starting point for defining the non-abelian T-dual of the principal chiral model with respect to its $F_L$ symmetry, $f\to Uf$. In order to proceed, one integrates out $A_\mu$ to arrive at the non-abelian T-dual theory defined in terms of the Lagrange multiplier field $\nu$: 
\EQ{
S_\text{T-Dual}=-\frac{\kappa^2}{\pi}\int d^2x\Tr\big[\partial_+\nu\big(1+\Ad(\nu)\big)^{-1}\partial_-\nu\big]\ .
\label{tdt}
}
This defines a new sigma-model with a non-compact target-space. In the context of string theory, this lack of compactness can lead to difficulties in interpreting the T-dual geometry. Sfetsos \cite{Sfetsos:2013wia} has proposed that one way to solve these global issues is to define a sort of ``regularized" T-dual theory by replacing the Lagrange multiplier term in \eqref{apcm} involving $\nu$ by the gauged WZW action for a group field $\CF$:
\EQ{
-\frac{\kappa^2}{\pi}\int d^2x\text{Tr}\big(\nu F_{+-}\big)\longrightarrow S_\text{gWZW}[\CF,A_\mu]\ .
}
There is then a sense that as $k\to\infty$ we can expand the group-valued field around the identity
as $\CF=I+\kappa^2\nu/k+\cdots$ and recover the T-dual theory \eqref{tdt}. Note that the ``regularized theory" is precisely the deformed theory with action \eqref{dWZW}. The precise sense in which the principal chiral model is recovered in the limit $k\to\infty$, or $\lambda\to1$, with fixed $\kappa$, remains to be determined. 

The other interesting limit involves taking $\lambda\to0$, or $\kappa^2\to\infty$ with fixed $k$. In this limit, it makes sense to integrate out the gauge field $A_\mu$ in \eqref{dWZW}. What results is a deformation of an ordinary (non-gauged) WZW model
\EQ{
S_\text{def}[\CF,A_\mu]=S_\text{WZW}[\CF]+\frac{4\pi}{\kappa^2}\int d^2x\,\Tr\big(\hat\JJ_+\hat\JJ_-\big)+\cdots\ ,
}
where $\hat\JJ_\pm$ are the usual currents of the ordinary WZW model:
\EQ{
\hat\JJ_+=-\frac k{2\pi}\CF^{-1}\partial_+\CF\ ,\qquad\hat\JJ_-=\frac k{2\pi}\partial_-\CF\CF^{-1}\ .
}
This kind of current-current deformation of WZW models are bosonized non-abelian Thirring models  \cite{Karabali:1988sz}. Such deformations are marginally relevant from a renormalization group point-of-view.

\subsection{Integrability}

Although Sfetsos has already proved integrability of the deformed theory by showing that its equations-of-motion are the same as the original principal chiral model in terms of the current $\JI_\mu$ defined in \eqref{icr}, here we take a different and simpler view that will generalize straighforwardly to the symmetric space theories later.

In much of what follows it will be useful to define the usual chiral currents of the F/F gauged WZW model,\EQ{
\JJ_+&=-\frac k{2\pi}\big(\CF ^{-1}\partial_+\CF +\CF ^{-1}A_+\CF -A_-\big)\ ,\\
\JJ_-&=\frac k{2\pi}\big(\partial_-\CF \CF ^{-1}-\CF A_-\CF ^{-1}+A_+\big)
\label{ksq2}
}
whose Poisson brackets take the form of two commuting classical Kac-Moody algebras
\EQ{
\big\{\JJ^a_\pm(x),\JJ^b_\pm(y)\big\}&=f^{abc}\JJ_\pm^c(y)\delta(x-y)\pm\frac {k}{2\pi}\delta^{ab}\delta'(x-y)\ ,\\
\big\{\JJ^a_+(x),\JJ^b_-(y)\big\}&=0\ ,\\
\label{km4}
}
Since the action does not depend on derivatives of $A_\mu$, 
the equations-of-motion of the gauge field $A_\mu$ are the constraints 
\EQ{
\JJ_\pm=-\frac k{2\pi}\Big(\frac1{\lambda }A_\pm-A_\mp\Big)\ .
\label{sd1}
}
Note that we will continue to refer to $A_\mu$ as a gauge field even though strictly speaking it is not one because the deformed theory \eqref{dWZW} is not gauge invariant. 
Note that when $\lambda^2\neq1$, the constraints above can be used to eliminate the gauge field in favour of the chiral currents:
\EQ{
A_\pm=-\frac{2\pi\lambda }{k(1-\lambda^2)}\big(\JJ_\pm+\lambda \JJ_\mp\big)\ .
\label{uu8}
}
Later we will see that these constraints are second class in the Hamiltonian formalism and so can be imposed strongly on the phase space at the expense of introducing Dirac brackets.

The equation-of-motion of the group field $\CF$ can be written either as
\EQ{
\big[\partial_++\CF ^{-1}\partial_+\CF +\CF ^{-1}A_+\CF ,\partial_-+A_-\big]=0\ ,
\label{sd2}
}
or, equivalently, by conjugating with $\CF$, as
\EQ{
\big[\partial_++A_+,\partial_--\partial_-\CF \CF ^{-1}+\CF  A_-\CF ^{-1}\big]=0\ .
\label{sd3}
}
Then, using \eqref{sd1} in \eqref{sd2} and \eqref{sd3}, leaves us with the pair of equations
\EQ{
&-\partial_-A_++\lambda \partial_+A_-+[A_+,A_-]=0\ ,\\
&-\lambda \partial_-A_++\partial_+A_-+[A_+,A_-]=0\ ,
\label{pls}
}
from which we find (for $\lambda\neq1$)
\EQ{
\partial_\mp A_\pm=\pm\frac1{1+\lambda }[A_+,A_-]\ .
}

If we define the current
\EQ{
\JI_\pm=\frac2{1+\lambda }A_\pm\equiv
-\frac{4\pi\lambda }{k(1+\lambda )(1-\lambda^2)}\big(\JJ_\pm+\lambda \JJ_\mp\big)\ ,
\label{pv2}
}
noting along the way the agreement with \eqref{gww},
then it is simple to see that these satisfy the equations-of-motion \eqref{h45}
which are equivalent to the equations-of-motion \eqref{em1} and Cartan-Mauer condition \eqref{cm1} of the principle chiral model. This proves the integrability of the deformed theory at the classical level.

Another more direct and satisfying way to proving integrability is via the Lax equation. The idea is to demand that the pair of equations \eqref{sd2} and \eqref{sd3}, together with the constraints \eqref{sd1}, are equivalent to the Lax equation \eqref{leq2} with
\EQ{
{\mathscr L}_\pm(z)=\frac{z}{z\pm1}\JI_\pm\,,\qquad \JI_\pm=\frac2{1+\lambda }A_\pm\,.
}
First of all, notice that for $z\not=0,\pm1$ the Lax equation \eqref{leq2} can be written as
\EQ{
\partial_+ J_- +\partial_-J_+ + z(\partial_+ J_- -\partial_-J_+ +[J_+,J_-])=0\,,
}
which exhibits that the equations-of-motion~\eqref{em1} and~\eqref{cm1}
follow from the Lax equation \eqref{leq2} if it is satisfied just for two distinct values $z=z_+$ and $z_-$.  
Then, we will identify \eqref{sd2} as the Lax equation for an specific choice of spectral parameter $z=z_+$ and similarly for \eqref{sd3} with $z=z_-$. This leads to four conditions,
\EQ{
-\frac{2\pi}k\JJ_\pm+A_\mp=\LL_\pm(z_\pm)\ ,\qquad A_\pm=\LL_\pm(z_\mp)\,,
}
which imply
\EQ{
\JJ_\pm= -\frac{k}{2\pi}\left(\frac{z_\pm(z_\mp \pm 1)}{z_\mp(z_\pm \pm1)}\,  A_\pm - A_\mp\right)\,.
}
The two parameters $z_\pm$  
are fixed by imposing the constraints \eqref{sd1} that relate $A_\mu$ to $\JJ_\mu$
\EQ{
\frac{z_\pm(z_\mp \pm 1)}{z_\mp(z_\pm \pm1)}= \lambda^{-1}\,,
}
which leads to
\EQ{
z_\pm=\pm\frac{\lambda+1}{\lambda-1}\,.
}
The conditions that $z_+\not= z_-$, and that they are $\not=0,\pm1$ and finite, requires that $\lambda^2\not=0,1$. All this is noteworthy because it shows how rigid the demand of integrability is and how perfectly it fits the deformed theory: integrability leaves little wriggle room.

\section{Hamiltonian Structure of the Deformed Principal Chiral Model}\label{s3}

In this section, we will investigate the Hamiltonian structure of the deformed principal chiral model with the goal of showing that 
it gives rise to the family of Poisson brackets \eqref{gt1}. 

First of all,  it should be no surprise that there is a very direct relation between the Kac-Moody current $\widetilde{\JJ}_\pm$ in \eqref{ksq} and the chiral current $\JJ_\pm$ of the WZW model.
In fact, taking into account the definition of $J_\mu$ in terms of $A_\mu$ and the constraints~\eqref{sd1}, one can check that $\widetilde{\JJ}_\pm =\JJ_\pm$
if we fix the normalization factor in~\eqref{ksq} to be
\EQ{
{\cal N}=\frac{(1+\lambda)^3}{8\lambda}\,.
}
Moreover, $\hat{k}{\cal N}=k$. 
In the following, we will argue that the classical Kac-Moody algebra~\eqref{km4} is not modified by the deformation in~\eqref{dWZW}---the Dirac brackets are equal to the Poisson brackets---and, therefore, the corresponding Poisson brackets coincide with the deformed Poisson brackets~\eqref{gt1} up to an overall ($\lambda$-dependent) normalization factor
\EQ{
\{\;,\;\}_{\hat{x}}= {\cal N} \{\;,\;\}= \frac{(1+\lambda)^3}{8\lambda} \{\;,\;\}\,.
\label{norm}
}
Since the equations-of-motion written in terms of $J_\pm$ are the same, this implies that the Hamiltonian of the deformed theory is
\EQ{
H={\cal N}H_{\text{PCM}}=-\frac{k}{16\pi} \frac{(1+\lambda)^3(1-\lambda)}{\lambda^2} \int dx\, \Tr(J_+^2+J_-^2)\,.
\label{hamnew}
}
Later in this section we will deduce this expression for $H$ using the Hamiltonian formalism.

Since Hamiltonian analysis is a key part of the current work, it is worth recalling some of the main features of Dirac's theory of constrained systems as it applies to field theories \cite{Dirac}. Suppose we have a Hamiltonian system with a set of constraints $\chi_i\approx0$. 
Dirac distinguishes between first class and second class constraints where the former have the property that their Poisson bracket with any other constraint vanishes
\EQ{
\{\chi_i,\chi_j\}\approx0\ ,\qquad\forall j\ ,\qquad i=\text{1st class}\ .
}
The $\approx0$ here, indicates that we can impose the constraints after the Poisson bracket has been evaluated. 
First class constraints correspond to local gauge symmetries. These should ultimately be fixed by introducing a gauge fixing condition one for each first class constraint. The full set of constraints, including the gauge fixing conditions are then effectively second class meaning more precisely that the inverse of the matrix
$C_{ij}=\{\chi_i,\chi_j\}$
exists. In that case, the second class constraints can be imposed ``strongly" on the phase space (so before Poisson brackets are evaluated) as long as the original Poisson bracket is replaced by the Dirac bracket:
\EQ{
\{{F},{G}\}^*= \{{F},{G}\}-\{{F},\chi_i\}C^{-1}_{ij}\{\chi_j,{G}\}\,.
\label{ddb}
}
Note that the physical dimension of the phase space is equal to
\EQ{
\text{dim phase space}=\EuScript N-2n_1-n_2\ ,
}
where $\EuScript N$ is the dimension of the original phase space before applying constraints, and $n_1$ and $n_2$ are the number of first and original second class constraints, respectively. 

One of the most useful properties of Dirac's theory is that one can apply the algorithm to find Dirac brackets iteratively. So even before gauge fixing the first class constraints, one can take any subset of second class constraints and impose them strongly on the phase space at the expense of introducing an intermediate Dirac bracket as above. One can then identify further subsets of second class constraints using them to define a further intermediate Dirac bracket. This can repeated as many times as one likes. In particular, there is no need to actually gauge fix as long as one is content to be left with first class constraints that reflect the unfixed gauge symmetry.

In general one usually has to worry about whether the set of constraints is complete in the sense that whether the requirement that a constraint is preserved in time leads to further constraints. In the present context, where we have a 
Lagrangian formalism, this complication does not arise since all constraints arise as Lagrange equations.

\vspace{0.2cm}\noindent
{\it The protection mechanism\/}:---In the following, we often use the fact that for certain quantities, the Dirac brackets are equal to the original Poisson brackets. 
The mechanism is very simple and is worth explaining. Suppose that we have a pair of subsets of second class constraints $\phi_p$ and $\psi_\alpha$ of equal size, with the property that the only non-vanishing Poisson brackets involving the constraints $\psi_p$ are 
$\{\phi_p,\psi_\alpha\}$ so that the matrix of Poisson brackets takes the block form
\EQ{
C_{ij}=\begin{blockarray}{cccc} 
&\psi_\alpha & \phi_p & \chi'_i\\
\begin{block}{c(ccc)}
\psi_\alpha & 0 & * & 0\\
\phi_p & * & * & *\\
\chi'_i & 0 & * & *\\
\end{block}
\end{blockarray}\qquad\implies\qquad
C_{ij}^{-1}=\begin{blockarray}{cccc} 
&\psi_\alpha & \phi_p & \chi'_i\\
\begin{block}{c(ccc)}
\psi_\alpha & * & * & *\\
\phi_p & * & 0 & 0\\
\chi'_i & * & 0 & *\\
\end{block}
\end{blockarray}\ .
}
where $\chi'_i$ are the remaining second class constraints.
So the inverse elements $C^{-1}_{pq}$, $C^{-1}_{pi}$ and $C^{-1}_{ip}$ vanish.

Consider now any quantities, say $F$ and $G$, that only have non-vanishing Poisson brackets with the subset $\phi_p$. Then, the existence of the constraints $\psi_q$ ``protects" the Poisson brackets $\{F,G\}$ from being altered by the Dirac procedure; namely,
\EQ{
\{F,G\}^*&=\{F,G\}-\{F,\chi_i\}C_{ij}^{-1}\{\chi_j,G\}\ ,\\
&=\{F,G\}-\{F,\phi_p\}C_{pq}^{-1}\{\phi_q,G\}=\{F,G\}\ .
}

In the following, we will draw heavily on the work of Bowcock \cite{Bowcock:1988xr} who considered the Hamiltonian structure of the gauged WZW model. Note, however, that our conventions are different in a number of respects. 

Now we turn to the deformed WZW model defined by the action \eqref{dWZW}.
The initial phase space is spanned by the current $\JJ_\mu$, the gauge field $A_\mu$ and the conjugate momenta to the gauge field $P_\mu$. The important point is that the Poisson brackets of the current $\JJ_\mu$ are just two decoupled classical Kac-Moody, or current, algebras as in \eqref{km4}. 
The components $\JJ_\pm$ then Poisson commute with the gauge field and its momentum. The latter have a standard Poisson bracket
\EQ{
\{P^a_\pm(x),A^b_\mp(y)\}=\frac12\delta^{ab}\delta(x-y)\ .
}

The Lagrangian does not depend on the time-derivative of the gauge field $A_\mu$ and so the conjugate momenta $P_\mu$ vanish. In the Hamiltonian formalism both the vanishing of $P_\mu$ and of its time-evolution, which provides the equations-of-motion of $A_\mu$~\eqref{sd1},
are viewed as constraints:
\EQ{
\chi_1&=P_+\approx0\ ,\qquad\chi_2=P_-\approx0\ ,\\
\chi_3&=\JJ_++\frac k{2\pi}\Big(\frac1{\lambda } A_+-A_-\Big)\approx0\ ,\\
\chi_4&=\JJ_-+\frac k{2\pi}\Big(\frac1{\lambda } A_--A_+\Big)\approx0\ .
}

Given our discussion of Dirac's theory, the key issue is the decomposition of the constraints into first and second class. It is this decomposition that will distinguish the deformed from the undeformed theory. The issue is settled by evaluating the matrix of the constraints
\EQ{
C_{ij}(x,y)=\{\chi_i(x),\chi_j(y)\}\approx\frac k{2\pi}\left(\begin{array}{cccc}0 & 0 & \,-1\, & \,\lambda^{-1}\, \\
0 & 0 & \,\lambda^{-1}\, & \,-1\, \\
\,1\, & \,-\lambda^{-1}\, & \,-D_-\, & 0 \\
\,-\lambda^{-1}\, & \,1\, & 0 & \,D_+\,\end{array}\right)\delta(x-y)
}
where we have defined the operators
\EQ{
D_\pm= \partial_x\pm\frac{2\pi}k \Ad(\JJ_\pm)\ .
\label{ddf}
}
In the above, and in much of the following, the Lie algebra indices have been left implicit.

In the undeformed theory $\lambda=1$ there are two sets of first class constraints. Clearly this includes $\chi_1+\chi_2$ but also (modulo second class constraints) $\chi_3+\chi_4$. However, in the deformed theory, $\lambda\neq1$, all the constraints are second class. This is entirely reasonable because, as we have already remarked, the deformed theory has no gauge symmetry; indeed, when $\lambda^2\neq1$ the matrix $C_{ij}(x,y)$ is invertible. Whilst the expression for the inverse is quite complicated, what is important is that its components in the $i=3,4$ subspace all vanish as can be seen by looking at the co-factors of these elements.

In the deformed theory, $\lambda^2\neq1$,
since all the constraints are second class,  we can set all of them strongly to zero by replacing the Poisson brackets by Dirac brackets \eqref{ddb}.
The simplest way to do this is to use the constraints to eliminate $P_\mu$ and $A_\mu$ from the physical phase space, i.e.~$A_\mu$ is given in terms of the current $\JJ_\mu$ by \eqref{uu8}, and then one should appreciate that, since $\{\chi_i(x),\chi_j(y)\}^{-1}$ vanishes in the $i=3,4$, subspace the Dirac brackets of $\JJ_\pm$ are the same as their original Poisson brackets \eqref{km4}. 
This is an example of the protection mechanism where the existence of the subset of constraints $\chi_1$ and $\chi_2$ which only have non-trivial Poisson brackets with $\chi_3$ and $\chi_4$ protects the Poisson brackets of $\JJ_\pm$ from being modified.

We have succeeded in showing that the deformed theories generalize the $\SU(2)$ principal chiral model summarized in section \ref{s2} to an arbitrary group $F$. In particular, the deformed Poisson brackets can be reformulated as two commuting classical Kac-Moody algebras.

The Hamiltonian on the physical phase space takes either of the forms
\EQ{
H&=-\frac{\pi}{k(1-\lambda^2)}\int dx\,\Tr\Big[(1+\lambda^2)(\JJ_+\JJ_+ +\JJ_-\JJ_-)+4\lambda \JJ_+\JJ_-\Big]\\[5pt]
&=-\frac{k(1-\lambda^2)}{4\pi\lambda^2}\int dx\, \Tr\big[A_+A_+ + A_- A_-\big]\,.
}
Given the expression of the current $\JI_\mu$ in \eqref{pv2}, we can also write the Hamiltonian as~\eqref{hamnew}

The conclusion is that one can view the deformation in two ways. Either we focus on $J_\mu$, in which case the  equations-of-motion are fixed while the symplectic structure changes, or we focus on $\JJ_\mu$, in which case the symplectic structure is fixed while the equations-of-motion change.

\section{The Deformed Symmetric Space Theories}\label{s4}

In the remainder of the paper, we take what we have learned about the deformed principal chiral model and find a way to  extend the same ideas to sigma-models with a symmetric space target-space $F/G$. These sigma-models are also known to be integrable.

We can define these sigma-models by a gauging procedure. That is we write a sigma-model for an $F$-valued field $f(x,t)$ and then gauge the subgroup $G\subset F$ which acts by right-multiplication $f\to fU$, $U\in G$. To this end we introduce a $\mathfrak g$-valued gauge field $B_\mu$ and write
\EQ{
S[f,B_\mu]=-\frac{\kappa^2}{\pi}\int d^2x\,\text{Tr}\big(\JI_+\JI_-\big)\ ,
\label{n3b}
}
where $\JI_\mu=f^{-1}\partial_\mu f-B_\mu$. The theory is invariant under a global $F$ left action $f\to Uf$, $U\in F$, in addition to the gauge symmetry acting to the right
\EQ{
f\to fU\ ,\qquad B_\mu\to U^{-1}\partial_\mu U+U^{-1}B_\mu U\ .
}

Symmetric spaces are special quotients of Lie groups $F/G$. They are associated to a particular $\mathbb Z_2$ automorphism of $\mathfrak f$ under which $\mf=\mf^{(0)}\oplus\mf^{(1)}$, where $\mf^{(0)}\equiv\mg$ is the Lie algebra of the subgroup $G\subset F$. We will denote a decomposition of any element of $A\in\mf$ as $A^{(0)}+A^{(1)}$. The Lie algebra $\mf$ respects the $\mathbb Z_2$ grading:
\EQ{
[\mf^{(i)},\mf^{(j)}]\subset f^{(i+j\ \text{mod}\,2)}\ .
}

Returning to the sigma model \eqref{n3b}, the equation-of-motion of the gauge field $B_\mu$ simply imposes the condition
\EQ{
\JI_\mu^{(0)}=0\qquad\implies\qquad B_\mu=(f^{-1}\partial_\mu f)^{(0)}\ .
}
The equation-of-motion of the group-valued field $f$ can be decomposed according to $\mg\oplus\mf^{(1)}$ as
\EQ{
&\partial_\pm \JI^{(1)}_\mp+[B_\pm,\JI^{(1)}_\mp]=0\ ,\\
&\partial_+B_--\partial_-B_++[B_+,B_-]+[J_+^{(1)},J_-^{(1)}]=0\ .
\label{eom16}
}

Classical integrability follows from writing these equations in terms of a Lax pair
\EQ{
[\partial_++{\mathscr L}_+(z),\partial_-+{\mathscr L}_-(z)]=0\
\label{leq3}
}
where 
\EQ{
{\mathscr L}_\pm(z)=B_\pm+z^{\pm1}\JI_\pm^{(1)}\ ,
}
where $z$ is the spectral parameter.

\subsection{The deformed theory}

We now follow essentially the same logic as for the principal chiral model to define a deformed theory. So we take an $F/G$ sigma-model and add to it a WZW model for an $F$-valued field $\CF$. Then one gauges the common $F$ action as in \eqref{fact} by introducing an $\mf$-valued gauge field $A_\mu$. So there are two gauge fields in play $A_\mu$ and $B_\mu$, the former valued in $\mf$ and the latter in $\mg$.

As previously we can fix the gauge symmetry by taking $f=I$, however, this only partially fixes the gauge symmetry from $F$ to $G$. The gauge field $B_\mu$ can now be trivially integrated out. This amounts to the replacement  
\EQ{
 -\frac{\kappa^2}{\pi} \int \Tr\big((A_+-B_+)(A_--B_-)\big)\longrightarrow  -\frac{\kappa^2}{\pi} \int \Tr\big(A_+^{(1)}A_-^{(1)}\big)\ .
}
Therefore the resulting theory takes the form of a deformation of an 
$F/F$ gauged WZW theory with action
\EQ{
S_\text{def}[\CF ,A_\mu]=S_\text{gWZW}[\CF ,A_\mu]-\frac k{\pi}\Big(\frac1{\lambda}-1\Big)\int d^2x\,\Tr\,\big(
A_+^{(1)}A_-^{(1)}\big)\ .
\label{dgWZW}
}
Since the deformation only involves $A^{(1)}_\mu$ the gauge symmetry is not completely broken and the subgroup $G$ remains intact.

As for the principal chiral model case, we expect that there are two interesting limits. Firstly, the limit $\lambda\to1$, or $k\to\infty$, at fixed $\kappa$, where we recover the non-abelian T-dual of the $F/G$ sigma-model with respect to its $F_L$ isometry. The other interesting limit is $\lambda\to0$, or $\kappa\to\infty$, with fixed $k$. In this limit, by integrating out the components $A^{(1)}_\mu$, the deformed theory can be interpreted as a current-current deformation of the a gauged $F/G$ WZW model
\EQ{
S_\text{def}[\CF,A_\mu]=S_\text{gWZW}[\CF,A_\mu^{(0)}]+\frac{4\pi}{\kappa^2}\int d^2x\,\text{Tr}\,\big(\hat\JJ_+^{(1)}\hat\JJ_-^{(1)}\big)+\cdots\ .
}
In the above, $\hat\JJ_\pm$ are the currents of the $F/G$ gauged WZW model
\EQ{
\hat\JJ_+&=-\frac k{2\pi}\big(\CF^{-1}\partial_+\CF+\CF^{-1}A_+^{(0)}\CF-A_-^{(0)}\big)\ ,\\
\hat\JJ_-&=\frac k{2\pi}\big(\partial_-\CF\CF^{-1}-\CF A_-^{(0)}\CF^{-1}+A_+^{(0)}\big)\ .
}

\subsection{Equations-of-motion and integrability}

If we define the usual chiral currents of the $F/F$ gauged WZW model as in 
\eqref{ksq2}
then the equations-of-motion of the gauge field take the form of constraints
\EQ{
\JJ_\pm=\mp\frac k{2\pi}A_1-\frac k{2\pi}\Big(\frac1{\lambda }-1\Big)A_\pm^{(1)}\ ,
\label{sd11}
}
where $\lambda$ is defined as in \eqref{ldf}.

The components of the constraints \eqref{sd11} valued in  $\mathfrak f^{(1)}$ (which later will be seen to be second class in the Hamilton-Dirac formalism) can be used to eliminate the gauge field components valued in $\mathfrak f^{(1)}$ in favour of the currents:
\EQ{
A_\pm^{(1)}=-\frac{2\pi\lambda }{k(1-\lambda^2)}\big(\JJ_\pm^{(1)}+\lambda \JJ_\mp^{(1)}\big)\ .
\label{uu9}
}

The equation-of-motion of the group field can be written as 
\EQ{
\big[\partial_++\CF ^{-1}\partial_+\CF +\CF ^{-1}A_+\CF ,\partial_-+A_-]=0
\label{eom5}
}
or equivalently, by conjugation with $\CF$, as
\EQ{
\big[\partial_--\partial_-\CF \CF ^{-1}+\CF A_-\CF ^{-1},\partial_++A_+]=0\ .
\label{eom6}
}

Using the constraints \eqref{sd11}, we can take the equations-of-motion \eqref{eom5} and \eqref{eom6} and write
them in terms of the gauge field.
Projecting onto $\mathfrak f^{(0)}\equiv\mathfrak g$ and $\mathfrak f^{(1)}$, we have the pair of equations
\EQ{
&-\partial_-A_+^{(1)}+\lambda \partial_+A_-^{(1)}+\lambda [A_+^{(0)},A_-^{(1)}]+[A_+^{(1)},A_-^{(0)}]=0\ ,\\
&-\lambda \partial_-A_+^{(1)}+\partial_+A_-^{(1)}+[A_+^{(0)},A_-^{(1)}]+\lambda [A_+^{(1)},A_-^{(0)}]=0\ ,
\label{j23}
}
along with 
\EQ{
-\partial_-A_+^{(0)}+\partial_+A_-^{(0)}+[A_+^{(0)},A_-^{(0)}]+\lambda^{-1}[A_+^{(1)},A_-^{(1)}]=0\ .
\label{hy2}
}

For generic $\lambda$, namely $\lambda^2\not=1$, the pair \eqref{j23} are equivalent to 
\EQ{
\partial_\mp A^{(1)}_\pm=[A^{(1)}_\pm,A^{(0)}_\mp]\ .
\label{yqq}
}
Then, if we define the currents $B_\mu\in\mg$  and $\JI_\mu^{(1)}$ via
\EQ{
 B_\pm=A^{(0)}_\pm\ ,\qquad \JI^{(1)}_\pm=\frac1{\sqrt\lambda}A^{(1)}_\pm\equiv
-\frac{2\pi\sqrt\lambda}{k(1-\lambda^2)}\big(\JJ_\pm^{(1)}+\lambda \JJ_\mp^{(1)}\big)\ ,
\label{rel2}
}
they satisfy precisely the same equations-of-motion \eqref{eom16} as the sigma-model. This indirectly proves classical integrability of the deformed theory. 

However, we can also go directly to the Lax equation by repeating the logic that we followed previously for the principal chiral model. The idea is to identify the pair of equations \eqref{eom5} and \eqref{eom6} as the Lax equation \eqref{leq3} evaluated for particular distinct values of the spectral parameter $z=z_\pm$, respectively. This means that
\EQ{
\JJ_\pm=-\frac k{2\pi}\big(\LL_\pm(z_\pm)-A_\mp\big)\ ,\qquad A_\pm=\LL_\pm(z_\mp)\ ,
}
for some fixed $z_\pm$ with $z_+\neq z_-$. 
From the second pair of equations it follows that
\EQ{
B_\pm=A^{(0)}_\pm\,,\qquad J_\pm^{(1)}= z_\mp^{\mp1} A_\pm^{(1)}\,.
}
Then, from the first pair we get
\EQ{
\JJ^{(0)}_\pm&=-\frac k{2\pi}\big(A_\pm^{(0)}-A_\mp^{(0)}\big)\ ,\\
\JJ^{(1)}_\pm&=-\frac k{2\pi}\Big(z_\pm^{\pm1} J_\pm^{(1)}-A_\mp^{(1)}\Big)=-\frac k{2\pi}\Big(\frac{z_+}{z_-}A_\pm^{(1)}-A_\mp^{(1)}\Big)\ ,
}
which are precisely the constraints \eqref{sd11} if we choose $z_-/z_+=\lambda$. This shows that the pair of equations \eqref{eom5} and \eqref{eom6}, together with the constraints \eqref{sd11}, can be written as the Lax equation \eqref{leq3} evaluated for $z=z_\pm=\lambda^{\mp1/2}$, 
which reproduces the equations we had previously. Moreover,
note that 
\EQ{
\LL_\pm(z)=A_\pm^{(0)}+z^{\pm1}\lambda^{-1/2}A_\pm^{(1)}\ .
}

The fact that the equations-of-motion of the deformed WZW theory are equal to the Lax equation \eqref{leq2} for two distinct values of $z$ is strong enough to prove equivalence to the Lax equation \eqref{leq2} which holds for arbitrary $z$. The reason why it is sufficient, is that the Lax equation has terms with powers $z^p$ ranging from $p=-1$ to $p=1$. However, for each projection of \eqref{leq2} on $\mf^{(i)}$ there are at most two independent equations, e.g.~for $\mf^{(1)}$ these are equations of order $z^{-1}$ and $z$. Consequently, it follows that if the Lax equation holds at two distinct values of $z$ then it holds for arbitrary $z$.

\section{The Hamiltonian Structure of the Deformed Theories}\label{s5}

In this section, we turn to the Hamiltonian structure of the deformed theories \eqref{dgWZW}. 

An important observation is that, since the deformation does not involve the field $\CF$, the chiral currents $\JJ_\pm$ defined in \eqref{ksq2} have the same Poisson brackets taking the form of two decoupled current algebras \eqref{km4}.
As in section \ref{s3}, there are four constraints arising from the vanishing of the canonical momenta to the gauge field and of their time-evolution, which are just the equations-of-motion of the gauge field,
\EQ{
&\chi_1= P_+\approx0\,,\qquad \chi_2= P_-\approx0\,,\\[5pt]
&\chi_3= {\JJ}_--\frac k{2\pi}A_1+\frac k{2\pi}\Big(\frac1{\lambda }-1\Big)A_+^{(1)}\approx0\,,\\[5pt]
&\chi_4= {\JJ}_++\frac k{2\pi}A_1+\frac k{2\pi}\Big(\frac1{\lambda }-1\Big)A_-^{(1)}\approx0\,.
\label{huu}
}
Note that, as previously, these constraints, being the Lagrangian equations-of-motion, are complete.

The situation now is like a hybrid of the deformed WZW discussed in section \ref{s3} and the gauged WZW according to the split $\mf=\mg\oplus\mf^{(1)}$.\footnote{In the following we will distinguish the generators in  ${\mathfrak g}$ and $ {\mathfrak f}^{(1)}$ using the following notation $\{T^A\}$ and $\{T^a\}$, respectively.}
 All the constraints in \eqref{huu} valued in $\mf^{(1)}$ behave like the constraints in section \ref{s3} and are second class. 
On the other hand the set of constraints valued in $\mg$ are partly first and partly second class to reflect the $G$ gauge symmetry of the deformed theory.

\vspace{0.3cm}
\noindent{\it Constraints in $\mf^{(1)}$}.---For these constraints we do not need to repeat the analysis of section \ref{s3}. The constraints can be used to eliminate $A_\pm^{(1)}$ in terms of $\JJ_\pm^{(1)}$ via
\EQ{
A_\pm^{(1)}=-\frac{2\pi\lambda}{k(1-\lambda^2)}\big(\JJ_\pm^{(1)}+ \lambda \JJ_\mp^{(1)}\big)\ .
}
Importantly, as in section \ref{s3}, the protection mechanism ensures that the Dirac brackets of $\JJ^{(1)}_\pm$ are just equal to the original Poisson brackets because the subset $(\chi_1^{(1)},\chi_2^{(2)})$ protects $(\chi_3^{(1)},\chi_4^{(1)})$.

\vspace{0.3cm}
\noindent{\it Constraints in $\mg$}.---For this set, the situation is exactly as in the gauged WZW model considered by Bowcock \cite{Bowcock:1988xr}. One can choose
\EQ{
P_1^{(0)}\approx0\ ,\qquad \JJ_-^{(0)}-\frac k{2\pi}A_1^{(0)}\approx0
}
as second class constraints. These can then be imposed strongly on the phase space to eliminate $P_1^{(0)}$ and $A_1^{(0)}$. Once again, the protection mechanism operates, so that the constraint $P_1^{(0)}$ protects the Poisson brackets of $\JJ_\pm^{(0)}$ from being modified.
This leaves two first class constraints,
\EQ{
P_0^{(0)}\approx0\ ,\qquad \JJ_+^{(0)}+\JJ_-^{(0)}\approx0\ .
}
which ultimately would have to be converted to second class by gauge fixing.

Let us summarize the situation that has emerged. Before gauge fixing the phase space is parameterized by $\JJ_\pm$ whose Dirac brackets are equal to their original Poisson brackets \eqref{km4}. We can count the dimension of the physical phase space by counting the first and second class constraints:

\begin{center}
\begin{tabular}{lcc}
\toprule
&first class&second class\\
\otoprule
$P_\mu^{(0)}$ & $\DG$ & $\DG$\\
$P_\mu^{(1)}$ & $0$ & $2\DFf$\\
$\JJ^{(0)}_\pm\pm\frac k{2\pi}A_1^{(0)}$ & $\DG$ & $\DG$\\
$\JJ^{(1)}_\pm+\frac k{2\pi}\big(\lambda^{-1}A_\pm^{(1)}-A_\mp^{(1)}\big)$ & $0$ & $2\DFf$\\
\bottomrule
\end{tabular}

\vspace{0.2cm}
$\boxed{\text{dim.~phase space}=6\DF-2n_1-n_2=2\DFf}$
\end{center}
Note that the final dimension is exactly what we expect of a theory whose configuration space is a deformation of the coset $F/G$.

\subsection{The Hamiltonian}

Dirac's theory of constrained systems also explains how to construct the Hamiltonian of the reduced theory that generates time translations with respect to the Dirac bracket. In the present case, the Hamiltonian before gauge fixing takes the form\footnote{Notice that $\dot{A}_0^{(0)}$ is arbitrary. The gauge transformations generated by $P_0^{(0)}\approx0$ can be fixed by imposing $A_0^{(0)}\approx0$ and, hence, $\dot{A}_0^{(0)}\approx0$.}
\EQ{
H&=-\frac{\pi}{k}\int dx\Tr\Big[\frac{1+\lambda^2}{1-\lambda^2} \left((\JJ_+^{(1)})^2+(\JJ_-^{(1)})^2\right) +\frac{4\lambda}{1-\lambda^2}\JJ_+^{(1)}\JJ_-^{(1)}\\[5pt] 
&\hspace{3.5cm}
-\frac{k}{\pi}A_0^{(0)} (\JJ_+^{(0)}+\JJ_-^{(0)}) +k \dot{A}_0^{(0)}P_0^{(0)}
\Big]\\[5pt]
&
\approx-\frac{\pi}{k(1-\lambda^2)}\int dx\, \Tr\Big[(\JJ_+^{(1)}+ \lambda\JJ_-^{(1)})^2 +(\JJ_-^{(1)}+ \lambda\JJ_+^{(1)})^2\Big]\,,
\label{changeH2}
}

The non-vanishing components of the energy-momentum tensor are
\EQ{
T_{\pm\pm}\approx -\frac\pi{k(1-\lambda^2)}\, \Tr\big(\JJ_\pm^{(1)}+ \lambda\JJ_\mp^{(1)}\big)^2\ ,
\label{emt}
}
so that $H=\int T_{00}\,dx= \int(T_{++}+T_{--})dx$. 
Equivalently, we can write
\EQ{
T_{\pm\pm}\approx-\frac{k(1-\lambda^2)}{4\pi\lambda}\, \Tr\big(\JI_\pm^{(1)}\big)^2\ .
\label{emt2}
}

\subsection{Relation to other work}

In this section we consider the relation of the Poisson (Dirac) brackets that we have derived and the Poisson brackets found for the deformation of the symmetric space sigma-models derived by Delduc et al.~in \cite{Delduc:2013fga}.

The Poisson brackets of Delduc et al~written in Appendix D of \cite{Delduc:2013fga} are defined in terms of phase space fields $\Pi$ and $\AA$ valued in $\mf$. Writing them explicitly in terms of components
\EQ{
\{\AA^{A}(x),\AA^{B}(y)\}'&=-\epsilon^2f^{ABC}(2\AA^{C}(y)+\Pi^{C}(y))\delta(x-y)+2\epsilon^2\delta^{AB}\delta'(x-y)\ ,\\
\{\AA^{A}(x),\AA^{b}(y)\}'&=-\epsilon^2f^{Abc}(\AA^{c}(y)+\Pi^{c}(y))\delta(x-y)\ ,\\
\{\AA^{a}(x),\AA^{b}(y)\}'&=-\epsilon^2f^{abC} \Pi^{C}(y)\delta(x-y)\ ,\\
\{\AA^{A}(x),\Pi^{B}(y)\}'&=f^{ABC} \AA^{C}(y)\delta(x-y)-\delta^{AB}\delta'(x-y)\ ,\\
\{\AA^{A}(x),\Pi^{b}(y)\}'&=(1-\epsilon^2)f^{Abc} \AA^{c}(y)\delta(x-y)-\epsilon^2 f^{Abc}\Pi^{c}(y)\delta(x-y)\ ,\\
\{\AA^{a}(x),\Pi^{B}(y)\}'&=f^{aBc} \AA^{c}(y)\delta(x-y)\ ,\\
\{\AA^{a}(x),\Pi^{b}(y)\}'&=f^{abC} \AA^{C}(y)\delta(x-y)+\epsilon^2 f^{abC}\Pi^{C}(y)\delta(x-y)-\delta^{ab}\delta'(x-y)\ ,\\
\{\Pi^{A}(x),\Pi^{B}(y)\}'&=f^{ABC}\Pi^{C}(y)\delta(x-y)\ ,\\
\{\Pi^{A}(x),\Pi^{b}(y)\}'&=f^{Abc} \Pi^{c}(y)\delta(x-y)\ ,\\
\{\Pi^{a}(x),\Pi^{b}(y)\}'&=(1-\epsilon^2)f^{abC}\Pi^{C}(y)\delta(x-y)\ ,
\label{p44}
}
These are written before applying the first class constraint $\Pi^{(0)}\approx0$.

For our deformation the Poisson (Dirac) brackets are just those of two decoupled classical Kac-Moody algebras \eqref{km4}. If we define the dictionary
\EQ{
\Pi^{(0)}&=-\frac{4\pi\lambda }{k(1-\lambda^2)}\big(\JJ_+^{(0)}+\JJ_-^{(0)}\big)\ ,~\qquad
\Pi^{(1)}=-\frac{2\pi\sqrt\lambda}{k(1-\lambda )}\big(\JJ_+^{(1)}+\JJ_-^{(1)}\big)\ ,\\
\AA^{(0)}&=-\frac{2\pi}{k(1+\lambda )}\big(\JJ_+^{(0)}-\lambda \JJ_-^{(0)}\big)\ ,\qquad
\AA^{(1)}=-\frac{2\pi\sqrt\lambda}{k(1+\lambda )}\big(\JJ_+^{(1)}-\JJ_-^{(1)}\big)\ .
}
or equivalently
\EQ{
\JJ_+&=-\frac k{4\pi}\big((1-\lambda )\Pi^{(0)}+2\AA^{(0)}\big)
-\frac k{4\pi\sqrt\lambda}\big((1-\lambda )\Pi^{(1)}+(1+\lambda )\AA^{(1)}\big)\ ,\\ 
\JJ_-&=\frac k{4\pi}\big((1-\lambda^{-1})\Pi^{(0)}+2\AA^{(0)}\big)
+\frac k{4\pi\sqrt\lambda}\big(-(1-\lambda )\Pi^{(1)}+(1+\lambda )\AA^{(1)}\big)
}
the Poisson brackets \eqref{p44} follow from \eqref{km4} with
\EQ{
\{F,G\}'=\frac{k(\lambda^2-1)}{4\pi\lambda }\{F,G\}\ ,\qquad
\epsilon^2=-\frac{(1-\lambda )^2}{4\lambda }\ .
\label{gt2}
}
In addition, the first class constraint
$\Pi^{(0)}\approx0$ is equivalent to the first class constraint $\JJ_+^{(0)}+\JJ_-^{(0)}\approx0$.

Notice that our deformation corresponds to the analytic continuation of the deformation parameter of \cite{Delduc:2013fga} to $\epsilon^2<0$. It is worth noticing that the Poisson brackets of  \cite{Delduc:2013fga} can also be written in terms of two decoupled classical Kac-Moody algebras for $\epsilon^2>1$. In contrast, for $0<\epsilon^2<1$, which is the case considered in that paper, the coefficients that relate the currents $\JJ_\pm$ to the fields $\Pi$ and $\AA$ become complex.

\section{Strings on Deformed Symmetric Spaces: the Pohlmeyer Reduction}\label{s6}

In this section, we consider the world-sheet theories that describe the classical motion of strings on a spacetime $\mathbb R\times{\mathfrak M}$, where $\mathbb R$ is the target-space time direction. In the gauge where the target-space time is proportional to the world-sheet time, the world-sheet theory involves a sigma-model on the spatial geometry $\mathfrak M$ with the additional Virasoro constraints 
\EQ{
T_{\pm\pm}=\text{constant}
\label{pohl}
} 
imposed, arising from the equations-of-motion of the world-sheet metric. In the context of an integrable sigma-model the resulting theory, the sigma-model plus Virasoro constraints and a partial gauge fixing, is known as the Pohlmeyer reduced theory \cite{Pohlmeyer:1975nb,Miramontes:2008wt}. In this section, we consider the Pohlmeyer reduction of the deformed symmetric space sigma-models constructed in section \ref{s4}. 

Given the components of the energy-momentum tensor in \eqref{emt} and~\eqref{emt2}, the Virasoro constraints can be written as 
\EQ{
T_{\pm\pm}\approx\frac{k \mu^2(1-\lambda^2)}{4\pi \lambda}
}
for a constant $\mu$ of unit mass dimension. The normalization here has been chosen so that the constraints take the form
\EQ{
\Tr\big(\JI_\pm^{(1)}\big)^2\approx-\mu^2\ .
}
This means that, at the level of the equations-of-motion written in terms of the current $\JI_\mu$, the Virasoro constraints do not depend on the deformation parameter $\lambda$.
In terms of the current $\JJ_\pm$, the constraints take the form
\EQ{
\Tr\big(\JJ_\pm^{(1)}+\lambda\JJ_\mp^{(1)})^2\approx-\frac{k^2\mu^2(1-\lambda^2)^2}{4\pi^2\lambda}\ .
\label{zll}
}
These constraints are second class since they satisfy
\EQ{
\{T_{\pm\pm}(x),T_{\pm\pm}(y)\}\approx
\frac{k\mu^2(1-\lambda^2)}{2\pi\lambda}\delta'(x-y)\ .
}
In addition, no new constraints are generated because $\partial_\mp T_{\pm\pm}=0$ and $T_{\pm\pm}\approx\,$const.~together imply that $\dot T_{\pm\pm}\approx0$. 

As well as imposing the Virasoro constraints, the Pohlmeyer reduction involves a partial gauge fixing. At the level of the equations-of-motion in the stringy context this has been explained by 
Tseytlin and Grigoriev \cite{Grigoriev:2007bu} (see also~\cite{Miramontes:2008wt}). What we will do now is to lift this analysis to the Poisson brackets.
The idea is that the solution to the Virasoro constraints breaks the gauge symmetry from $G$ to a subgroup $H$ 
and so it makes sense to fix the gauge symmetry from $G$ to leave the subgroup $H$. In order to do this, we
impose, instead of the $+$ constraint of \eqref{zll}, the more refined constraint
\EQ{
\varphi_2=\JJ_+^{(1)}+\lambda\JJ_-^{(1)}+\frac{k\mu(1-\lambda^2)}{2\pi\sqrt\lambda}\Lambda\approx0\ ,
\label{imm}
}
where $\Lambda$ is a constant element of $\mf^{(1)}$ with $\Tr(\Lambda^2)=-1$. 
For simplicity, we shall concentrate on rank 1 symmetric spaces for which $\Lambda$ is unique, up to conjugation in $G$. 

The unbroken $H$ gauge symmetry corresponds to the elements of $U\in G$ that stabilize $\Lambda$: $U\Lambda U^{-1}=\Lambda$. In what follows, we will need the further decomposition of $\mg$ into the image and the kernel of $\Lambda$:
\EQ{
{\mathfrak g}= {\mathfrak g}^\perp\oplus {\mathfrak g}^\parallel\,,\qquad
{\mathfrak g}^\perp= {\mathfrak g}\cap \text{Ker ad}(\Lambda)\,,\qquad
{\mathfrak g}^\parallel= {\mathfrak g}\cap \text{Im ad}(\Lambda)\ .
}
The kernel $\mg^\perp$ is identified with the Lie algebra $\mh$ of the subgroup $H$.

The time derivative of the constraint \eqref{imm} is 
\EQ{
\partial_0\varphi_2\approx \partial_1\varphi_2-\frac{k\mu(1-\lambda^2)}{2\pi\sqrt\lambda}[A_-^{(0)},\Lambda]\approx -\frac{k\mu(1-\lambda^2)}{2\pi\sqrt\lambda}[A_-^{(0)},\Lambda]\ ,
}
so in the partially gauge fixed theory there is a secondary constraint
\EQ{
\varphi_3=A_-^\parallel\approx0\ .
}

Our strategy for dealing with the new constraints is to firstly impose all the second class constraints from the original set $\chi_i$ in \eqref{huu} to leave the field coordinates on the intermediate phase space as $\JJ_\pm$, $A_0^{(0)}$ and $P_0^{(0)}$. The new Pohlmeyer constraints then take the form
\EQ{
\varphi_1&=\Tr\big(\JJ_-^{(1)} +\lambda \JJ_+^{(1)}\big)^2 +\frac{k^2\mu^2(1-\lambda^2)^2}{4\pi^2\lambda}\approx0\,,\\
\varphi_2&=\JJ_+^{(1)} +\lambda \JJ_-^{(1)}+\frac{k\mu(1-\lambda^2)}{2\pi\sqrt\lambda}\Lambda \approx 0\,,\\
\varphi_3&=\JJ_-^\parallel-\frac k{2\pi}A_0^\parallel\approx 0\,,
}
which have to be imposed, in addition to the former first class constraints (which we re-label)
\EQ{
\varphi_4=P_0^{(0)}\approx0\ ,\qquad\varphi_5=\JJ_+^{(0)}+\JJ_-^{(0)}\approx0\ .
}

It is clear that the subset of constraints $\varphi_4^\parallel$ and $\varphi_5^\parallel$, the ones valued in $\mg^\parallel$,
do not Poisson commute with $\varphi_3^\parallel$ and so become second class. This is a manifestation of the fact that the gauge symmetry is reduced from $G$ to $H$. The remaining first class constraints are
\EQ{
\varphi_4^\perp=P_0^\perp\approx0\ ,\qquad\varphi_5^\perp=\JJ_+^\perp+\JJ_-^\perp\approx0\ .
}

Exploiting once again the iterative property of the Dirac procedure, we can eliminate the second class constraints $\varphi_3$ and $\varphi_4^\parallel$. The matrix of constraints in the $(\varphi_3,\varphi_4^\parallel)$ subspace takes the form
\EQ{
C_{ij}(x,y)=-\frac {k}{2\pi}\left(\begin{array}{cc}D & -1\\ \,1~ &  0\end{array}\right)\delta(x-y)\ .
}
Here, $D\equiv D_-\approx D_+$ is defined as in \eqref{ddf} implicitly taken to act from 
$\mg^\parallel$ to $\mg^\parallel$. We can use the constraints to to set $P_0^\parallel=0$ and 
$A_0^\parallel=2\pi\JJ_-^\parallel/k$ strongly on the phase space.
From the form of the constraint matrix it follows that the Dirac brackets of $\JJ_-^\parallel$ are not affected: the constraint $\varphi_4^\parallel$ acts as a protection. Consequently at this stage $\JJ_\pm$ still have Dirac brackets that are equal to the initial Poisson brackets \eqref{km4}.

The remaining second class constraints are 
\EQ{
\varphi_1&=\Tr\big(\JJ_-^{(1)} +\lambda \JJ_+^{(1)}\big)^2 +\frac{k^2\mu^2(1-\lambda^2)^2}{4\pi^2\lambda}\approx0\,,\\
\varphi_2&=\JJ_+^{(1)} +\lambda \JJ_-^{(1)}+\frac{k\mu(1-\lambda^2)}{2\pi\sqrt\lambda}\Lambda\approx 0\,,\\
\varphi_5&=\JJ_+^\parallel+\JJ_-^\parallel\approx0\ .
}
The Poisson brackets of these remaining constraints take the form
\EQ{
&
\{\varphi_1,\varphi_1\}\approx \frac{k^3\mu^2(1-\lambda^2)^3}{2\pi^3\sqrt\lambda}\delta'(x-y)\,,\qquad
\{\varphi_1,\varphi_2\}\approx 0\,,\qquad
\{\varphi_1,\varphi_5\}\approx 0\,,\\
&
\{\varphi_2,\varphi_2\}\approx \frac k{2\pi}(1-\lambda^2)D\delta(x-y)\,,\qquad
\{\varphi_2,\varphi_5\}\approx \frac{k\mu(1-\lambda^2)}{2\pi\sqrt\lambda}\,\Ad(\Lambda)\delta(x-y)
\\[5pt]
&
\{\varphi_5,\varphi_5\}\approx 0\,.
}
They can be written in matrix form as follows ($i,j=1,2,5$)
\EQ{
C_{ij}(x,y)&=\{\varphi_i(x), \varphi_j(y)\} \\ &\approx \frac k{2\pi}\left(
\arraycolsep=1.4pt\def\arraystretch{2}
\begin{array}{ccc}
~\frac{k^2\mu^2(1-\lambda^2)^3}{\pi^2\sqrt\lambda}\partial_x~ & 0 & 0 \\
0 &~(1-\lambda^2) D & \frac{k\mu(1-\lambda^2)}{\pi\sqrt\lambda}\,\Ad(\Lambda)~\\
0 &~  \frac{k\mu(1-\lambda^2)}{\pi\sqrt\lambda}\,\Ad(\Lambda)~& 0\end{array}\right)\delta(x-y)\,.
}
In the above, in our index free notation, the operators $D$ and $\Ad(\Lambda)$ are restricted to act between the appropriate spaces.

For $\lambda^2\not=1$ all these constraints are second class and can be imposed strongly on the phase space.
In this case, the Dirac brackets are different from the original Poisson brackets of $\JJ_\pm$. The Dirac brackets can be extracted from the inverse constraint-matrix
\EQ{
&C_{ij}(x,y)^{-1}\\ &\approx\frac{2\pi}k\left(
\arraycolsep=1.4pt\def\arraystretch{2}
\begin{array}{ccc}
~-\frac{\pi^2\lambda}{k^2(1-\lambda^2)^3\mu^2} \partial_x^{-1}~ & 0 & 0 \\ 
0 & 0 & ~\frac{\pi\sqrt\lambda}{k\mu(1-\lambda^2)}\text{ad\,}(\Lambda)^{-1}~ \\ 
0  & ~\frac{\pi\sqrt\lambda}{k\mu(1-\lambda^2)}\text{ad\,}(\Lambda)^{-1}~ &~ \frac{\pi^2\lambda}{k^2\mu^2(1-\lambda^2)} \text{ad\,}(\Lambda)^{-1} D\,\text{ad\,}(\Lambda)^{-1}~\end{array}\right)\delta(x-y)\ .
\label{vgg}
}

Let us consider the Dirac bracket of $\JJ_-^{(0)}$ with itself. In order to write the result cleanly we will define the quantity $\tilde\Lambda$ via
\EQ{
\JJ_-^{(1)}+\lambda \JJ_+^{(1)}=-\frac{k\mu(1-\lambda^2)}{2\pi\sqrt\lambda}\tilde\Lambda\ ,
\label{b33}
}
mirroring the constraint $\varphi_2$, so that the Pohlmeyer constraint $\varphi_1$ implies 
\EQ{
\text{Tr}\,\tilde\Lambda^2\approx-1\ .
\label{pn4}
}
Then we can write the Dirac bracket in an index free way as
\EQ{
&\{\JJ_-(x),\JJ_-(y)\}^*=-\frac k{2\pi}D\delta(x-y)\\ &
+\frac{k\sqrt\lambda}{2\pi(1-\lambda^2)}\Big(\frac{1}{\mu^2}D\Ad(\Lambda)^{-1}D\Ad(\Lambda)^{-1}D-\mu^2[\tilde\Lambda,\Lambda]\partial_x^{-1}[\Lambda,\tilde\Lambda]
\Big)\delta(x-y)\ .
\label{ka1}
}
Other Dirac brackets can be calculated as needed.

It would be interesting to relate these Dirac brackets to the deformed Poisson brackets defined directly from the underlying integrable hierarchy on the light front in \cite{Hollowood:2014fha,Mikhailov:2006uc}.

The dimension of the physical phase space of the Pohlmeyer reduced theory is obtained by counting the number of first and second class constraints. The result is $2(\DG-\DH)$ as summarized in the table:

\begin{center}
\begin{tabular}{lcc}
\toprule
&first class&second class\\
\otoprule
$P_\mu^{(0)}$ & $\DG$ & $\DG$\\
$P_\mu^{(1)}$ & $0$ & $2\DFf$\\
$\JJ^{(0)}_\pm\pm\frac k{2\pi}A_1^{(0)}$ & $\DG$ & $\DG$\\
$\JJ^{(1)}_\pm+\frac k{2\pi}\big(\lambda^{-1}A_\pm^{(1)}-A_\mp^{(1)}\big)$ & $0$ & $2\DFf$\\
$T_{\pm\pm}-\frac{k\mu^2(1-\lambda^2)}{4\pi\lambda}$ & 0 & 2\\
\bottomrule
\end{tabular}

\vspace{0.2cm}
$\boxed{\text{dim.~phase space}=2\DFf-2=2(\DG-\DH)}$
\end{center}

The counting above can be re-done in the partially gauge fixed formalism. Of course the counting must be the same.
In the partially gauge fixed theory, there are $1+\DFf+(\DG-\DH)$ additional second class constraints but then in addition $2(\DG-\DH)$ original first constraints become second class. The new counting is summarized in the table below.
\begin{center}
\begin{tabular}{lcc}
\toprule
&first class&second class\\
\otoprule
$P_\mu^{(0)}$ & $\DH$ & $2\DG-\DH$\\
$P_\mu^{(1)}$ & $0$ & $2\DFf$\\
$\JJ^{(0)}_\pm\pm\frac k{2\pi}A_1^{(0)}$ & $\DH$ & $2\DG-\DH$\\
$\JJ^{(1)}_\pm+\frac k{2\pi}\big(\lambda^{-1}A_\pm^{(1)}-A_\mp^{(1)}\big)$ & $0$ & $2\DFf$\\
$\Tr\big(\JJ_-^{(1)}+\lambda \JJ_+^{(1)}\big)^2+\frac{k^2\mu^2(1-\lambda^2)^2}{4\pi^2\lambda }$ & 0 & 1\\
$\JJ_+^{(1)}+\lambda \JJ_-^{(1)}+\frac{k\mu(1-\lambda^2)}{2\pi\sqrt\lambda}\Lambda$ & 0 & $\DFf$\\
$A_-^{\parallel}$ & 0 & $\DG-\DH$\\
\bottomrule
\end{tabular}

\vspace{0.2cm}
$\boxed{\text{dim. phase space}=2(\DG-\DH)}$
\end{center}

\subsection{A parameterization}

The Pohlmeyer constraint $\varphi_1$ can be solved explicitly by introducing a parameterization of the form
\EQ{
\tilde\Lambda=\gamma^{-1}\Lambda\gamma\ ,
\label{b44}
}
in terms of a $G$-valued field $\gamma$.
The equation-of-motion \eqref{yqq} for the lower sign becomes
\EQ{
\partial_+\big(\gamma^{-1}\Lambda\gamma\big)=\big[\gamma^{-1}\Lambda\gamma,A_+^{(0)}\big]\ ,
}
which implies that in terms of the $\gamma$ parameterization 
\EQ{
A_+^{(0)}=\gamma^{-1}\partial_+\gamma+\gamma^{-1}W_+\gamma\ ,
\label{pu8}
}
where $[W_+,\Lambda]=0$ so $W_+\in\mh$. If we denote $A_-^{(0)}=W_-$, then the constraint $\varphi_3$ implies that $[W_-,\Lambda]=0$ so $W_-\in\mh$. Using the constraints, we then find
the relation
\EQ{
\JJ_-^{(0)}=\frac k{2\pi}\big(A_+^{(0)}-A_-^{(0)}\big)=\frac k{2\pi}\big(\gamma^{-1}\partial_+\gamma+\gamma^{-1}W_+\gamma-W_-\big)\ .
\label{pgt}
}
In addition, the equation-of-motion \eqref{hy2} on the constraint surface becomes
\EQ{
\big[\partial_++\gamma^{-1}\partial_+\gamma+\gamma^{-1}W_+\gamma+\mu\Lambda,
\partial_-+W_-+\mu\gamma^{-1}\Lambda\gamma\big]=0
\label{emr}
}
which is independent of the deformation parameter $\lambda$. This equation is known as the symmetric space sine-Gordon (SSSG) equation. In an on-shell  gauge where $W_\mu=0$ they are known as non-abelian affine Toda equations and as the name suggests they can be viewed as a non-abelian generalization of the affine Toda equations \cite{Hollowood:1994vx,Ferreira:1994rc,FernandezPousa:1996hi,Ferreira:1996sm,Nirov:2006nv}.

\section{The Symmetric Space Sine-Gordon Theories}\label{s7}

We will find that the deformed string world-sheet theory has a very interesting 
limit as $\lambda\to0$. In order to set the scene, let us first describe a class of relativistic integrable field theories associated to a symmetric space $F/G$ that generalize the sine-Gordon theory \cite{Park:1994bx,Hollowood:1994vx,Bakas:1995bm}.

These theories have an action which takes the form of a gauged WZW model for the coset $G/H$, which is {\it not\/} to be confused with the symmetric space $F/G$, deformed by a potential term:
\EQ{
S_\text{SSSG}=S_\text{gWZW}[\gamma,W_\mu]-\frac {k\mu^2}{\pi}\int d^2x\,\Tr\left(\Lambda
\gamma^{-1}\Lambda\gamma\right)\ .
\label{ala}
}
Here, $S_\text{gWZW}[\gamma,W_\mu]$ is the usual gauged WZW action for $G/H$ written for instance in \eqref{gWZW}. Note that the potential term must be written as a trace in the larger Lie algebra $\mf$. This means that the equation-of-motion is naturally written as an equation in $\mf$:
\EQ{
\big[\partial_++\gamma^{-1}\partial_+\gamma+\gamma^{-1}W_+\gamma+\mu\Lambda,
\partial_-+W_-+\mu\gamma^{-1}\Lambda\gamma\big]=0\ .
}
This is precisely \eqref{emr}. The equations-of-motion of the $\mh$-valued gauge field $W_\mu$ are the usual constraints
\EQ{
{\cal J}_\pm^\perp\approx\mp\frac k{2\pi}W_1\ ,
}
where we have defined the usual chiral current
\EQ{
{\cal J}_+&=-\frac k{2\pi}\big(\gamma^{-1}\partial_+\gamma +\gamma ^{-1}W_+\gamma -W_-\big)\ ,\\
{\cal J}_-&=\frac k{2\pi}\big(\partial_-\gamma\gamma^{-1}-\gamma W_-\gamma^{-1}+W_+\big)
\label{kmc2}
}
that takes values in ${\mathfrak g}$.

The Hamiltonian structure of the theory follows from our previous discussions. Before gauge fixing, the phase space is spanned by ${\cal J}_\pm$, $W_0$ and its conjugate momentum ${\cal P}_0$. There are then first class constraints
\EQ{
{\cal P}_0\approx0\qquad\text{and}\qquad{\cal J}_+^\perp+{\cal J}_-^\perp\approx0
}
that reflect the $H$ gauge symmetry.

Note that instead of taking the phase space coordinate fields ${\cal J}_\pm$ we can equally well take, say, ${\cal J}_+$ and $\gamma$, the $G$-valued field itself. These fields have Poisson brackets
\EQ{
\{{\cal J}_+^A(x),{\cal J}_+^B(y)\}_\text{SSSG}&=-f^{ABC}{\cal J}_+^C(y)\delta(x-y)-\frac k{2\pi}\delta^{AB}\delta'(x-y)\ ,\\
\{{\cal J}_+^A(x),\gamma(y)\}_\text{SSSG}&=\gamma(y) T^A\delta(x-y)\ ,\\
\{\gamma(x),\gamma(y)\}_\text{SSSG}&=0\ .\phantom{\frac k2}
\label{pbsg}
}
Notice that relative to our earlier choice of convention \eqref{km4}, there are some different sign conventions in the brackets above. This difference is physically irrelevant but is useful for what will come in the following section.

\subsection{Making the link with the deformed theories}

Going back to our deformed $F/F$ WZW theory and its Pohlmeyer reduction and, more specifically, to its Hamiltonian structure investigated in section \ref{s6}, we see that in the limit $\lambda\to0$ the inverse matrix of constraints \eqref{vgg} vanishes. This means that the Dirac brackets of $\JJ_\pm$ are equal to their original Poisson brackets. In particular, making the indices explicit, the Dirac bracket \eqref{ka1} becomes simply
\EQ{
\{\JJ_-^A(x),\JJ_-^B(y)\}^*=f^{ABC}\JJ_-^C(y)\delta(x-y)-\frac k{2\pi}\delta'(x-y)\ .
\label{ka2}
}
In other words a classical current algebra for the subgroup $G$.
In addition, in the same limit we have 
\EQ{
\{\JJ_-^A(x),(\gamma^{-1}\Lambda\gamma)^b(y)\}^*&=f^{Abc}(\gamma^{-1}\Lambda\gamma)^c(y)\delta(x-y)\ ,\\
\{(\gamma^{-1}\Lambda\gamma)^a(x),(\gamma^{-1}\Lambda\gamma)^b(y)\}^*&=0\ ,\phantom{\frac k2}
\label{ka3}
}

If we reflect on the Dirac brackets \eqref{ka2}, \eqref{ka3} and \eqref{pbsg} then a striking new interpretation presents itself. The fields $\gamma$ and $W_\mu$ of the sine-Gordon theory are identified with the fields of the same notation in the 
the deformed WZW model. In particular, $\gamma$ is valued in $G\subset F$ and $W_\mu$ is an $\mh$-valued connection. This SSSG theory has a gauged WZW current \eqref{kmc2} with
\EQ{
{\cal J}_+=-\frac k{2\pi}\big(\gamma^{-1}\partial_+\gamma+\gamma^{-1}W_+\gamma-W_-\big)\ .
\label{kbb}
}
that becomes identified with $-\JJ_-^{(0)}$. Recalling from \eqref{pu8} and the discussion below, that
\EQ{
A_+^{(0)}=\gamma^{-1}\partial_+\gamma+\gamma^{-1}W_+\gamma\ ,\qquad A_-^{(0)}=W_-\ ,
}
means that  the gauge field $W_\pm$ defined in section \ref{s6} is identified precisely with the gauge field $W_\pm$ defined in section \ref{s7}. As a check, the constraint for $\JJ_-^{(0)}$ in \eqref{huu} is precisely \eqref{kbb}. Hence, the complete
relation between the original variables and the auxiliary ones in the limit $\lambda\to0$ is
\EQ{
\JJ_-^{(0)}&=-{\cal J}_+\ ,\qquad A^{(0)}_+=\gamma^{-1}\partial_+\gamma+\gamma^{-1}W_+\gamma\ ,\qquad A_-^{(0)}=W_-\ ,\\ P_0^\perp&={\cal P}_0
\qquad~~~\JJ_-^{(1)}=-\frac{k\mu}{2\pi\sqrt\lambda}\gamma^{-1}\Lambda\gamma\ ,\qquad
\JJ_+^{(1)}=-\frac{k\mu}{2\pi\sqrt\lambda}\Lambda
}
the last two relations following from \eqref{b33} and \eqref{b44}. The Poisson/Dirac brackets \eqref{ka3} are then entirely consistent with those of the SSSG theory \eqref{pbsg} with $\{F,G\}_\text{SSSG}=\{F,G\}^*$. Note that the dimension of the physical phase space of the Pohlmeyer reduced theory was shown to be $2(\DG-\DH)$ precisely the physical dimension of the phase space of the SSSG theory.

This proves that at the classical level, the Pohlmeyer reduction of our deformed theory, the theory that describes the string world-sheet, in the limit $\lambda\to0$ is a relativistic generalization of the sine-Gordon theory.

The deformed Poisson brackets in the limit $\lambda\to0$ and the relation with the symmetric space sine-Gordon theory have also been investigated in \cite{Delduc:2012qb,Delduc:2012vq} from a different perspective of alleviating the non-ultra-locality of the $\lambda\to1$ Poisson bracket.

\section{Discussion}\label{s8}

It has always been intriguing that the world-sheet sigma-model in the gauge-gravity correspondence has an equation-of-motion that can be written in a relativistic form \cite{Pohlmeyer:1975nb,Grigoriev:2007bu,Mikhailov:2007xr,Grigoriev:2008jq,Miramontes:2008wt,Hollowood:2009tw,Hollowood:2009sc,Hollowood:2010dt,Hollowood:2011fq}. This observation led to a lot of activity leading to an S-matrix that interpolates between the stringy S-matrix and the S-matrix of a relativistic QFT \cite{Hoare:2011wr,Hoare:2013ysa}. In another direction, this related relativistic theory was shown to have a symplectic structure that does not suffer from the non-ultra-locality of the string theory symplectic structure \cite{Delduc:2012qb,Delduc:2012vq}. In this work we have shown how to construct a series of deformations of the string sigma-model that in a certain limit yield the relativistic sine-Gordon theory directly at the Lagrangian level.

\subsection{The quantum theories}

Naturally, one is interested in the deformed theories at the quantum level. If we turn to the principal chiral model case and, specifically for $\SU(2)$, then we can make some definite statements based on \cite{Evans:1994hi}. It is worth recalling the main points of that work.  The principal chiral model is asymptotically free meaning that the coupling $\kappa$ undergoes renormalisation group flow running as $\kappa\to\infty$, or $\lambda\to0$, in the UV. 
It is known that integrability survives in the quantum theory and therefore the theory has an exact factorizable $S$-matrix. The whole $S$-matrix is built out of the 2 particle to 2-particle elements that have the schematic form
\EQ{
S(\theta)=\sigma(\theta)R_F(\theta)\otimes R_F(\theta)\ ,
}
where $\theta=\theta_1-\theta_2$ is the rapidity difference of the incoming states and $\sigma(\theta)$ is a scalar factor: the dressing phase. Each of the factors $R_F(\theta)$ carries group indices and accounts for the non-trivial exchange of quantum numbers. 
These are associated to the so-called rational solutions of the Yang-Baxter equation that are associated to the Yangian of $F$.
The fact that there are two factors of this type reflects the $F_L\times F_R$ symmetry of the theory.

The work \cite{Evans:1994hi} shows that in the deformed theory the coupling $\kappa$ still runs. However, the UV limit is no longer free; rather, as we saw in section \ref{s2}, when $\kappa\to\infty$ the UV theory is an $\SU(2)$ WZW model. Such a theory is ``asymptotically CFT" rather than ``asymptotically free" \cite{Evans:1994hi}. The S-matrix of the deformed theory then takes the form
\EQ{
S(\theta)=\sigma(\theta)R_{U_q(\SU(2))}(\theta)\otimes R_{\SU(2)}(\theta)\ ,
}
where the $\SU(2)_L$ factor, the one that is gauged in order to construct the deformed theory, has changed from the rational
$\SU(2)$ factor to the quantum group generalization where the quantum  group deformation parameter is $q=-e^{-i\pi/(k+2)}$. In addition, this quantum group solution of the Yang-Baxter equation is taken in the so-called interaction-round-a-face" (IRF) picture, or ``restricted-solid-on-solid" (RSOS) picture. The significance of this is that the quantum numbers in this sector are naturally interpretated as kinks: see figure \ref{f1}. This is to be contrasted with the original $R_F$ factor which appears in the ``vertex" picture. 

\FIG{
\begin{tikzpicture} [line width=1.5pt,inner sep=2mm,
place/.style={circle,draw=blue!50,fill=blue!20,thick}]
\begin{pgfonlayer}{foreground layer}
\node at (1.5,1.5) [place] (sm) {$S$}; 
\end{pgfonlayer}
\node at (0,0) (i1) {$i$};
\node at (3,0) (i2) {$j$};
\node at (0,3) (i3) {$k$};
\node at (3,3) (i4) {$l$};
\draw[->] (i1) -- (i4);
\draw[->] (i2) -- (i3);
\begin{scope}[xshift=6cm]
\begin{pgfonlayer}{foreground layer}
\node at (1.5,1.5) [place] (sm) {$S$}; 
\end{pgfonlayer}
\node at (0,0) (i1) {};
\node at (3,0) (i2) {};
\node at (0,3) (i3) {};
\node at (3,3) (i4) {};
\node at (0,1.5) (j1) {$a$};
\node at (1.5,0) (j2) {$b$};
\node at (3,1.5) (j3) {$c$};
\node at (1.5,3) (j4) {$d$};
\draw[->] (i1) -- (i4);
\draw[->] (i2) -- (i3);
\end{scope}
\end{tikzpicture}
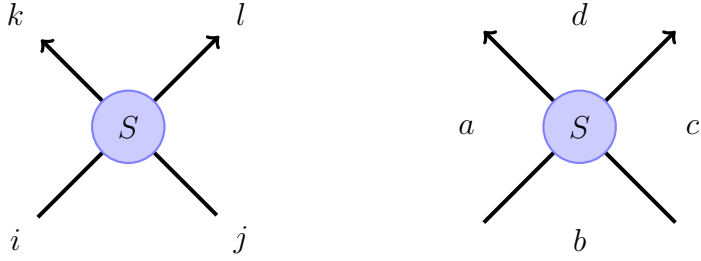
\caption{\small The vertex and IRF labels for the 2-body S-matrix elements. For the case of $\SU(2)$ in the vertex picture, the particles are labelled by $i,j,k,l\in\{\pm\frac12\}$  the weights of the spin $\frac12$ representation of $\SU(2)$. In the IRF or SOS picture on the right, one labels the vacua  $a,b,c,d\in\{0,\frac12,1,\frac32,\ldots\}$ which are spins of arbitrary irreducible representations of $\SU(2)$ with $|a-b|=|b-c|=|a-d|=|d-c|=\frac12$. The particles are then interpretated as kinks $K_{ab}+K_{bc}\to K_{ad}+K_{dc}$ interpolating between adjacent vacua. In the restricted case the spins are restricted to the finite set $\{0,\frac12,1,\ldots,\frac k2-1\}$, where $q=e^{i\pi/k}$.
\label{f1}
}}

In the limit $k\to\infty$, the RSOS formulation becomes unrestricted and it would be fascinating to understand the relation between the scattering theories with the vertex $R_F$ and IRF $R_{U_q(F)}$ (with $k=\infty$, i.e.~$q=1$) and non-abelian T-duality.

This picture generalises from $\SU(2)$ to arbitrary $F$.  The one-loop running of the coupling 
confirms that, as for $\SU(2)$, $\lambda\to0$ in the UV \cite{Itsios:2014lca,Sfetsos:2014jfa}.

For the symmetric space sigma-models the $S$-matrices take the form
\EQ{
S(\theta)=\sigma(\theta)R_F(\theta)\ ,
}
reflecting the $F_L$ isometry of the sigma-model. We conjecture that the S-matrices of the deformed theory will take the form
\EQ{
S(\theta)=\sigma(\theta)R_{U_q(F)}(\theta)\ .
}
Interestingly, we can view the principal chiral models as symmetric space sigma-models themselves for $G\times G/G$. So  
one can conjecture that in this case the deformed theories will involve a gauged WZW model for $G/G\times G/G$ and so there can be two independent levels $k_1$ and $k_2$. For these theories, we conjecture that the exact S-matrices are then
\EQ{
S(\theta)=\sigma(\theta)R_{U_{q_1}(F)}(\theta)\otimes R_{U_{q_2}(F)}(\theta)\ ,
}
with $q_i=-e^{-i\pi/(k_i+c_2(F))}$.

\subsection{Semi-symmetric spaces}

The case of most interest to string theory is when we consider the world-sheet theory of the string moving in AdS$_5\times S^5$. The world-sheet theory in the Green-Schwarz formalism takes the form of an integrable sigma-model on a semi-symmetric space $\text{PSU}(2,2|4)/\text{Sp}(2,2)\times\text{Sp}(4)$. Notice that the group $F$ in the numerator of the coset is now a super Lie group. This sigma-model is known to be integrable \cite{Bena:2003wd}.

The question is whether there is a deformation of this world-sheet theory of the kind that generalizes the ones we have constructed for ordinary symmetric spaces. This question will be investigated more fully elsewhere \cite{us}, but we can at least write down the action of the deformed theory. As in \eqref{dWZW} it takes the form of the a deformation of an $F/F$ gauged WZW model:
\EQ{
S_\text{def}[\CF ,A_\mu]=S_\text{gWZW}[\CF ,A_\mu]-\frac k{\pi}\int d^2x\,\STr\Big[
A_+\big(\Omega-1\big)A_-\Big]\ .
\label{dWZW2}
}
The group $F$ is now the supergroup $\text{PSU}(2,2|4)$ and traces are replaced by super-traces. The deformation is defined by the constant matrix $\Omega$ and it will be shown that for a very specific choice of $\Omega$, namely\footnote{Note that we could have written \eqref{dgWZW} in the same way with $\Omega=\mathbb P^{(0)}+\lambda^{-1}\mathbb P^{(1)}$ in that case.}
\EQ{
\Omega=\mathbb P^{(0)}+\frac1{\lambda}{\mathbb P}^{(1)}+\frac1{\lambda^2}{\mathbb P}^{(2)}+\lambda\mathbb P^{(3)}\ ,
}
where ${\mathbb P}^{(i)}$ are the projectors on the Lie super algebra $\mf$ associated to the eigenspaces $\mf^{(i)}$ of the $\mathbb Z_4$ automorphism associated to the semi-symmetric space, the resulting theory (i) has the same equation-of-motion as the original Green-Schwarz sigma-model, (ii) is classically integrable and (iii) has fermionic gauge (kappa) symmetries which are crucial to getting a consistent world-sheet theory with the correct number of fermionic degrees-of-freedom.

The conjecture is that $\lambda$ is an exactly marginal coupling and so the deformed theory is a consistent string background. It is then very natural to suggest that the S-matrix constructed in \cite{Hoare:2011wr} is precisely the S-matrix of the string world-sheet theory. This S-matrix depends on two couplings $g$ and $q$. The latter takes specific values $q=e^{i\pi/k}$ for integer $k$ which naturally identified it as level of the gauged WZW model in \eqref{dWZW2}. The other coupling $g$ is some function of the deformation parameter $\lambda$. One important fact about the S-matrix in \cite{Hoare:2013ysa} is that it is written in IRF or RSOS form and this matches the discussion of the principal chiral model above.

It is interesting to note that the relation between the deformation parameter of the quantum group is $q=e^{i\pi/k}$ for the semi-symmetric space case and $q=-e^{-i\pi/(k+c_2(F))}$ for the symmetric space case. Of course $c_2(F)$, the quadratic Casimir of the adjoint representation, is known to vanish for the Lie supergroup $\text{PSU}(2,2|4)$. However the other sign differences have a dramatic effect on the $S$-matrices. In the symmetric space case, the bound states fall in  the anti-symmetric representations of the (quantum) group whereas in the semi-symmetric space case they fall in symmetric representations. The number of anti-symmetric representations only depends on the rank of $F$ whereas the number of symmetric representations only depends on the level $k$.

\vspace{0.2cm}
\begin{center}
{\tiny******}
\end{center}
\vspace{0.2cm}

\noindent We would like to thank Kostas Sfetsos for fruitful discussions. TJH is supported in part by the STFC grant ST/G000506/1. 
JLM is supported in part by MINECO (FPA2011-22594), the Spanish Consolider-
Ingenio 2010 Programme CPAN (CSD2007-00042), Xunta de Galicia (GRC2013-024),
and FEDER.
DMS is supported by the FAPESP-BEPE grant 2013/23328-1.




\begin{thebibliography}{99}

{\small



\bibitem{Beisert:2010jr}
  N.~Beisert, C.~Ahn, L.~F.~Alday, Z.~Bajnok, J.~M.~Drummond, L.~Freyhult, N.~Gromov and R.~A.~Janik {\em et al.},
  ``Review of AdS/CFT Integrability: An Overview,''
  Lett.\ Math.\ Phys.\  {\bf 99} (2012) 3
  [arXiv:1012.3982 [hep-th]].

\bibitem{Arutyunov:2009ga}
  G.~Arutyunov and S.~Frolov,
  ``Foundations of the $AdS_5 \times S^5$ Superstring. Part I,''
  J.\ Phys.\ A {\bf 42} (2009) 254003
  [arXiv:0901.4937 [hep-th]].

\bibitem{Serg}
V.V.~Serganova, ``Classification of real simple Lie superalgebras and symmetric superspaces,"  Funct. Anal. Appl. {\bf17} (1983) 200
 
\bibitem{Zarembo:2010sg}
  K.~Zarembo,
  ``Strings on Semisymmetric Superspaces,''
  JHEP {\bf 1005} (2010) 002
  [arXiv:1003.0465 [hep-th]].

\bibitem{Beisert:2008tw}
  N.~Beisert and P.~Koroteev,
  ``Quantum Deformations of the One-Dimensional Hubbard Model,''
  J.\ Phys.\ A {\bf 41} (2008) 255204
  [arXiv:0802.0777 [hep-th]].



\bibitem{Hoare:2011wr}
  B.~Hoare, T.~J.~Hollowood and J.~L.~Miramontes,
  ``q-Deformation of the $\text{AdS}_5 \times S^5$ Superstring S-matrix and its Relativistic Limit,''
  JHEP {\bf 1203} (2012) 015
  [arXiv:1112.4485 [hep-th]].


\bibitem{Hoare:2012fc}
  B.~Hoare, T.~J.~Hollowood and J.~L.~Miramontes,
  ``Bound States of the q-Deformed $AdS_5 x S^5$ Superstring S-matrix,''
  JHEP {\bf 1210} (2012) 076
  [arXiv:1206.0010 [hep-th]].
  
\bibitem{Hoare:2013ysa}
  B.~Hoare, T.~J.~Hollowood and J.~L.~Miramontes,
  ``Restoring Unitarity in the q-Deformed World-Sheet S-Matrix,''
  JHEP {\bf 1310} (2013) 050
  [arXiv:1303.1447 [hep-th]].

   
  


 
  
\bibitem{LeClair:1989wy}
  A.~LeClair,
  ``Restricted Sine-Gordon Theory and the Minimal Conformal Series,''
  Phys.\ Lett.\ B {\bf 230} (1989) 103.

\bibitem{Bernard:1990cw}
  D.~Bernard and A.~LeClair,
  ``Residual Quantum Symmetries Of The Restricted Sine-gordon Theories,''
  Nucl.\ Phys.\ B {\bf 340} (1990) 721.

\bibitem{Bernard:1990ys}
  D.~Bernard and A.~LeClair,
  ``Quantum group symmetries and nonlocal currents in 2-D QFT,''
  Commun.\ Math.\ Phys.\  {\bf 142} (1991) 99



\bibitem{Arutyunov:2012zt}
  G.~Arutyunov, M.~de Leeuw and S.~J.~van Tongeren,
  ``The Quantum Deformed Mirror TBA I,''
  JHEP {\bf 1210} (2012) 090
  [arXiv:1208.3478 [hep-th]].


\bibitem{Arutyunov:2012ai}
  G.~Arutyunov, M.~de Leeuw and S.~J.~van Tongeren,
  ``The Quantum Deformed Mirror TBA II,''
  JHEP {\bf 1302} (2013) 012
  [arXiv:1210.8185 [hep-th]].

 
\bibitem{Delduc:2013fga}
  F.~Delduc, M.~Magro and B.~Vicedo,
  ``On classical $q$-deformations of integrable sigma-models,''
  JHEP {\bf 1311} (2013) 192
  [arXiv:1308.3581 [hep-th]].


\bibitem{Delduc:2013qra}
  F.~Delduc, M.~Magro and B.~Vicedo,
  ``An integrable deformation of the $AdS_5 \times S^5$ superstring action,''
  Phys.\ Rev.\ Lett.\  {\bf 112} (2014) 051601
  [arXiv:1309.5850 [hep-th]].

\bibitem{Delduc:2014kha}
  F.~Delduc, M.~Magro and B.~Vicedo,
 ``Derivation of the action and symmetries of the q-deformed $AdS_5 x S^5$ superstring,''
  arXiv:1406.6286 [hep-th].

\bibitem{Kawaguchi:2014qwa}
  I.~Kawaguchi, T.~Matsumoto and K.~Yoshida,
  ``Jordanian deformations of the AdS$_5 \times S^5$ superstring,''
  JHEP {\bf 1404} (2014) 153
  [arXiv:1401.4855 [hep-th]].

\bibitem{Kawaguchi:2014fca}
  I.~Kawaguchi, T.~Matsumoto and K.~Yoshida,
  JHEP {\bf 1406} (2014) 146
  [arXiv:1402.6147 [hep-th]].
  
\bibitem{Klimcik:2002zj}
  C.~Klimcik,
  ``Yang-Baxter sigma-models and dS/AdS T duality,''
  JHEP {\bf 0212} (2002) 051
  [hep-th/0210095].

 
\bibitem{Arutyunov:2013ega}
  G.~Arutyunov, R.~Borsato and S.~Frolov,
  ``S-matrix for strings on $\eta$-deformed $AdS_{5} \times S^5$,''
  JHEP {\bf 1404} (2014) 002
  [arXiv:1312.3542 [hep-th]].

    
\bibitem{Arutynov:2014ota}
  G.~Arutyunov, M.~de Leeuw and S.~J.~van Tongeren,
  ``On the exact spectrum and mirror duality of the $(AdS_5 x S^5)_\eta$ superstring,''
  arXiv:1403.6104 [hep-th].

\bibitem{Arutyunov:2014cda}
  G.~Arutyunov and D.~Medina-Rinc\'on,
  ``Deformed Neumann model from spinning strings on $(AdS_5 x S^5)_\eta$,''
  arXiv:1406.2536 [hep-th].

\bibitem{Hoare:2014pna}
  B.~Hoare, R.~Roiban and A.~A.~Tseytlin,
  ``On deformations of AdS$_n\times S^n$ supercosets,''
  JHEP {\bf 1406} (2014) 002
  [arXiv:1403.5517 [hep-th]].
 
\bibitem{Kameyama:2014bua}
  T.~Kameyama and K.~Yoshida,
  ``Anisotropic Landau-Lifshitz sigma models from $q$-deformed AdS$_5\times S^5$ superstrings,''
  arXiv:1405.4467 [hep-th].
  
\bibitem{Hollowood:2014fha}
  T.~J.~Hollowood and J.~L.~Miramontes,
  ``Symplectic Deformations of Integrable Field Theories and AdS/CFT,''
  arXiv:1403.1899 [hep-th].

  
\bibitem{Pohlmeyer:1975nb}
  K.~Pohlmeyer,
  ``Integrable Hamiltonian Systems And Interactions Through Quadratic
  Commun.\ Math.\ Phys.\  {\bf 46} (1976) 207.
  
\bibitem{Grigoriev:2007bu}
  M.~Grigoriev and A.~A.~Tseytlin,
  ``Pohlmeyer reduction of $AdS_5 \times S^5$ superstring sigma-model,''
  Nucl.\ Phys.\ B {\bf 800} (2008) 450
  [arXiv:0711.0155 [hep-th]].

\bibitem{Mikhailov:2007xr}
  A.~Mikhailov and S.~Schafer-Nameki,
  ``Sine-Gordon-like action for the Superstring in $AdS_5 \times S^5$,''
  JHEP {\bf 0805} (2008) 075
  [arXiv:0711.0195 [hep-th]].


\bibitem{Grigoriev:2008jq}
  M.~Grigoriev and A.~A.~Tseytlin,
  Int.\ J.\ Mod.\ Phys.\ A {\bf 23} (2008) 2107
  [arXiv:0806.2623 [hep-th]].



\bibitem{Hollowood:2013oca}
  T.~J.~Hollowood, J.~L.~Miramontes and D.~M.~Schmidtt,
  ``The Structure of Non-Abelian Kinks,''
  JHEP {\bf 1310} (2013) 058
  [arXiv:1306.6651 [hep-th]].



\bibitem{Hollowood:2011fq}
  T.~J.~Hollowood and J.~L.~Miramontes,
  ``The $AdS_5 x S_5$ Semi-Symmetric Space Sine-Gordon Theory,''
  JHEP {\bf 1105} (2011) 136
  [arXiv:1104.2429 [hep-th]].
    
\bibitem{Hoare:2011nd}
  B.~Hoare, T.~J.~Hollowood and J.~L.~Miramontes,
  ``A Relativistic Relative of the Magnon S-Matrix,''
  JHEP {\bf 1111} (2011) 048
  [arXiv:1107.0628 [hep-th]].


\bibitem{Balog:1993es}
  J.~Balog, P.~Forgacs, Z.~Horvath and L.~Palla,
  ``A New family of SU(2) symmetric integrable sigma-models,''
  Phys.\ Lett.\ B {\bf 324} (1994) 403
  [hep-th/9307030].
  
\bibitem{Evans:1994hi}
  J.~M.~Evans and T.~J.~Hollowood,
  ``Integrable theories that are asymptotically CFT,''
  Nucl.\ Phys.\ B {\bf 438} (1995) 469
  [hep-th/9407113].

  
\bibitem{Sfetsos:2013wia}
  K.~Sfetsos,
  ``Integrable interpolations: From exact CFTs to non-Abelian T-duals,''
  Nucl.\ Phys.\ B {\bf 880} (2014) 225
  [arXiv:1312.4560 [hep-th]].

\bibitem{Miramontes:2008wt}
  J.~L.~Miramontes,
  ``Pohlmeyer reduction revisited,''
  JHEP {\bf 0810} (2008) 087
  [arXiv:0808.3365 [hep-th]].
 
    
\bibitem{Luscher:1977rq}
  M.~Luscher and K.~Pohlmeyer,
  ``Scattering of Massless Lumps and Nonlocal Charges in the Two-Dimensional Classical Nonlinear sigma-Model,''
  Nucl.\ Phys.\ B {\bf 137} (1978) 46.

\bibitem{Brezin:1979am}
  E.~Brezin, C.~Itzykson, J.~Zinn-Justin and J.~B.~Zuber,
  ``Remarks About the Existence of Nonlocal Charges in Two-Dimensional Models,''
  Phys.\ Lett.\ B {\bf 82} (1979) 442.

\bibitem{Bernard:1990jw}
  D.~Bernard,
  ``Hidden Yangians in 2-D massive current algebras,''
  Commun.\ Math.\ Phys.\  {\bf 137} (1991) 191.

\bibitem{Nappi:1979ig}
  C.~R.~Nappi,
  ``Some Properties of an Analog of the Nonlinear $\sigma$ Model,''
  Phys.\ Rev.\ D {\bf 21} (1980) 418.
  
\bibitem{Rajeev:1988hq}
  S.~G.~Rajeev,
  ``Nonabelian Bosonization Without Wess-zumino Terms. 1. New Current Algebra,''
  Phys.\ Lett.\ B {\bf 217} (1989) 123.
 
  

\bibitem{Bowcock:1988xr}
  P.~Bowcock,
  ``Canonical Quantization of the Gauged Wess-Zumino Model,''
  Nucl.\ Phys.\ B {\bf 316} (1989) 80.
 
\bibitem{Faddeev:1985qu}
  L.~D.~Faddeev and N.~Y.~.Reshetikhin,
  ``Integrability of the Principal Chiral Field Model in (1+1)-dimension,''
  Annals Phys.\  {\bf 167} (1986) 227.
  
\bibitem{Karabali:1988au}
  D.~Karabali, Q-H.~Park, H.~J.~Schnitzer and Z.~Yang,
  ``A GKO Construction Based on a Path Integral Formulation of Gauged Wess-Zumino-Witten Actions,''
  Phys.\ Lett.\ B {\bf 216} (1989) 307.

\bibitem{Gawedzki:1988hq}
  K.~Gawedzki and A.~Kupiainen,
  ``G/h Conformal Field Theory from Gauged WZW Model,''
  Phys.\ Lett.\ B {\bf 215} (1988) 119.
  
\bibitem{Karabali:1989dk}
  D.~Karabali and H.~J.~Schnitzer,
  ``BRST Quantization of the Gauged WZW Action and Coset Conformal Field Theories,''
  Nucl.\ Phys.\ B {\bf 329} (1990) 649.

\bibitem{Tseytlin:1993hm}
  A.~A.~Tseytlin,
  ``On A 'Universal' class of WZW type conformal models,''
  Nucl.\ Phys.\ B {\bf 418} (1994) 173
  [hep-th/9311062].

 
\bibitem{Sfetsos:1994vz}
  K.~Sfetsos,
  ``Gauged WZW models and nonAbelian duality,''
  Phys.\ Rev.\ D {\bf 50} (1994) 2784
  [hep-th/9402031].
  
\bibitem{Polychronakos:2010hd}
  A.~P.~Polychronakos and K.~Sfetsos,
  ``High spin limits and non-abelian T-duality,''
  Nucl.\ Phys.\ B {\bf 843} (2011) 344
  [arXiv:1008.3909 [hep-th]].
 
\bibitem{Karabali:1988sz}
  D.~Karabali, Q-H.~Park and H.~J.~Schnitzer,
  ``Thirring Interactions, Nonabelian Bose-fermi Equivalences and Conformal Invariance,''
  Nucl.\ Phys.\ B {\bf 323} (1989) 572.
    
\bibitem{Dirac} 
P.~A.~M.~Dirac, ``Generalized Hamiltonian dynamics". Canadian Journal of Mathematics 2 (1950) 129;   ``Lectures on quantum mechanics." Belfer Graduate School of Science Monographs Series 2 (1964) New York.  

\bibitem{Mikhailov:2006uc} 
  A.~Mikhailov,
  ``Bihamiltonian structure of the classical superstring in $AdS_5\times S^5$,''
  Adv.\ Theor.\ Math.\ Phys.\  {\bf 14}, 1585 (2010)
  [hep-th/0609108].
 
\bibitem{Hollowood:1994vx}
  T.~J.~Hollowood, J.~L.~Miramontes and Q-H.~Park,
  ``Massive integrable soliton theories,''
  Nucl.\ Phys.\ B {\bf 445} (1995) 451
  [hep-th/9412062].
 
\bibitem{Ferreira:1994rc}
  L.~A.~Ferreira, J.~L.~Miramontes and J.~Sanchez Guillen,
  ``Solitons, tau functions and Hamiltonian reduction for nonAbelian conformal affine Toda theories,''
  Nucl.\ Phys.\ B {\bf 449} (1995) 631
  [hep-th/9412127].

\bibitem{FernandezPousa:1996hi}
  C.~R.~Fernandez-Pousa, M.~V.~Gallas, T.~J.~Hollowood and J.~L.~Miramontes,
  ``The Symmetric space and homogeneous sine-Gordon theories,''
  Nucl.\ Phys.\ B {\bf 484} (1997) 609
  [hep-th/9606032].
 
\bibitem{Ferreira:1996sm}
  L.~A.~Ferreira, J.~L.~Miramontes and J.~Sanchez Guillen,
  ``Tau functions and dressing transformations for zero curvature affine integrable equations,''
  J.\ Math.\ Phys.\  {\bf 38} (1997) 882
  [hep-th/9606066].

\bibitem{Nirov:2006nv}
  K.~.S.~Nirov and A.~V.~Razumov,
  ``Toda equations associated with loop groups of complex classical Lie groups,''
  Nucl.\ Phys.\ B {\bf 782} (2007) 241
  [math-ph/0612054].
 
\bibitem{Bakas:1995bm}
  I.~Bakas, Q-H.~Park and H.~-J.~Shin,
  ``Lagrangian formulation of symmetric space sine-Gordon models,''
  Phys.\ Lett.\ B {\bf 372} (1996) 45
  [hep-th/9512030].
  
\bibitem{Park:1994bx}
  Q-H.~Park,
  ``Deformed coset models from gauged WZW actions,''
  Phys.\ Lett.\ B {\bf 328} (1994) 329
  [hep-th/9402038].

\bibitem{Delduc:2012qb}
  F.~Delduc, M.~Magro and B.~Vicedo,
  ``Alleviating the non-ultralocality of coset sigma-models through a generalized Faddeev-Reshetikhin procedure,''
  JHEP {\bf 1208} (2012) 019
  [arXiv:1204.0766 [hep-th]].
 
\bibitem{Delduc:2012vq}
  F.~Delduc, M.~Magro and B.~Vicedo,
  ``Alleviating the non-ultralocality of the $AdS_5 \times S^5$ superstring,''
  JHEP {\bf 1210} (2012) 061
  [arXiv:1206.6050 [hep-th]].
 
\bibitem{Hollowood:2009tw}
  T.~J.~Hollowood and J.~L.~Miramontes,
  ``Magnons, their Solitonic Avatars and the Pohlmeyer Reduction,''
  JHEP {\bf 0904} (2009) 060
  [arXiv:0902.2405 [hep-th]].

\bibitem{Hollowood:2009sc}
  T.~J.~Hollowood and J.~L.~Miramontes,
  ``A New and Elementary CP**n Dyonic Magnon,''
  JHEP {\bf 0908} (2009) 109
  [arXiv:0905.2534 [hep-th]].

\bibitem{Hollowood:2010dt}
  T.~J.~Hollowood and J.~L.~Miramontes,
  ``Classical and Quantum Solitons in the Symmetric Space Sine-Gordon Theories,''
  JHEP {\bf 1104} (2011) 119
  [arXiv:1012.0716 [hep-th]].
    
\bibitem{Sfetsos:2014jfa}
  K.~Sfetsos and K.~Siampos,
  ``Gauged WZW-type theories and the all-loop anisotropic non-Abelian Thirring model,''
  arXiv:1405.7803 [hep-th].
  
\bibitem{Itsios:2014lca}
  G.~Itsios, K.~Sfetsos and K.~Siampos,
  ``The all-loop non-Abelian Thirring model and its RG flow,''
  Phys.\ Lett.\ B {\bf 733} (2014) 265
  [arXiv:1404.3748 [hep-th]].

\bibitem{Bena:2003wd}
  I.~Bena, J.~Polchinski and R.~Roiban,
  ``Hidden symmetries of the AdS(5) x S**5 superstring,''
  Phys.\ Rev.\ D {\bf 69} (2004) 046002
  [hep-th/0305116].

\bibitem{us}
  T.~J.~Hollowood, J.~L.~Miramontes and D.~M.~Schmidtt,
{\it to appear\/}.


  
  

}

\end{thebibliography}
\end{document}